\documentclass{jfm}
\usepackage{graphicx}
\usepackage{overpic}
\usepackage{amsmath}
\usepackage{epstopdf, epsfig}
\usepackage{subfigure}
\usepackage{xcolor}
\usepackage[T1]{fontenc}
\definecolor{blue}{rgb}{0,0,1}

\usepackage{comment}
\usepackage{upgreek}
\usepackage{multirow}

\usepackage{float}
\usepackage{booktabs}

\usepackage[colorlinks=true, citecolor=blue, linkcolor=blue, urlcolor=blue]{hyperref}

\shortauthor{K.-J. Qian, Z.-J. Li et al.}

\title[Journal of Fluid Mechanics]{A radiation two-phase flow model for simulating plasma-liquid interactions}
\author{Ke-Jian Qian\aff{1}\footnotemark[1], Zhu-Jun Li\aff{1}\footnotemark[1], Tao Tao\aff{2,3}, De-Hua Zhang\aff{1}, \\ Rui Yan\aff{1,3}\footnotemark[2], and Hang Ding\aff{1}\footnotemark[2]
\footnotetext[1]{The authors contributed equally to the article.}
\footnotetext[2]{Email address: ruiyan@ustc.edu.cn (R. Yan), hding@ustc.edu.cn (H. Ding)}
 }
\affiliation{$^1$State Key Laboratory of High Temperature Gas Dynamics, School of Engineering Science, University of Science and Technology of China, Hefei 230026, China\\
$^2$Department of Plasma Physics and Fusion Engineering, University of Science and Technology of China, Hefei 230026, China \\
$^3$Joint Team for the Double-cone Ignition Scheme}
\begin{document}

\label{firstpage}
\maketitle

\begin{abstract}
In laser-produced plasma (LPP) extreme ultraviolet (EUV) sources,  deformation of a tin droplet into an optimal target shape is governed by its interaction with a pre-pulse laser-generated plasma. This interaction is mediated by a transient ablation pressure, whose complex spatio-temporal evolution remains experimentally inaccessible. Existing modeling approaches are limited: Empirical pressure-impulse models neglect dynamic plasma feedback, while advanced radiation-hydrodynamic codes often fail to resolve late-time droplet hydrodynamics. To bridge this gap, we propose a radiation two-phase flow model based on a diffuse interface methodology. The model integrates radiation hydrodynamics for the plasma with the Euler equations for a weakly compressible liquid, extending a five-equation diffuse interface formulation to incorporate radiation transport, thermal conduction, and ionization. This formulation enforces pressure and velocity equilibrium across the diffuse interface region, with closure models constructed to ensure correct jump conditions at interfaces and asymptotically recover the pure-phase equations in bulk regions. Then, we apply the model to simulate a benchmark pre-pulse scenario, where a $50\ \upmu \mathrm{m}$ tin droplet is irradiated by a $10\ \mathrm{ns}$ laser pulse. The simulations capture the rapid plasma expansion and subsequent inertial flattening of the droplet into a thin, curved sheet over microsecond timescales. Notably, the model reproduces experimentally observed features such as an axial jet—rarely replicated in prior simulations. Quantitative agreement with experimental data for sheet dimensions and velocity validates the approach. The proposed model self-consistently couples laser-plasma physics with compressible droplet dynamics, providing a powerful tool for fundamental studies of plasma-liquid interactions in LPP-EUV source optimization.
\end{abstract}


\section{Introduction}
\label{sec:intro}
Laser-produced plasmas (LPP) are the sources of extreme ultraviolet (EUV) light for nanolithography~\citep{versolato2019}. Generation of EUV light in a modern LPP-EUV device involves two steps: a relatively low intensity pre-pulse laser is to prepare the target by deforming a tin droplet into a proper tin sheet; then a subsequent main pulse heats it up to produce EUV-emitting plasmas~\citep{mizoguchi2010, van2020}. To maximize the conversion efficiency and minimize tin debris, a precise control of droplet shape induced by the pre-pulse is crucial~\citep{versolato2022} and requires in-depth understanding on the droplet-plasma dynamics~\citep{meijer2022}. Upon pre-pulse impact, asymmetric plasma expansion from localized laser energy absorption exerts very high pressure at the droplet's illuminated surface~\citep{basko2015, kurilovich2018, sheil2024}. This pressure, usually referred to as the ablation pressure, not only launches acoustic waves traveling inside the droplet~\citep{reijers2017}, but also propels the droplet and simultaneously deforms it into a thin sheet which eventually fragments~\citep{gelderblom2016}. The ablation pressure represents the interaction between the plasma and the droplet, and its spatial distribution significantly changes with time. However, the ablation pressure profile remains experimentally unmeasurable. 

To numerically investigate droplet dynamics, different ablation pressure models have been proposed to mimic the impact from the plasma~\citep{klein2015, gelderblom2016, francca2025}. In the numerical simulations, the droplet was assumed to be incompressible and inviscid. \citet{gelderblom2016} simulated deformation of a droplet after being impacted by a laser pulse using a boundary integral method coupled with a simplified ablation pressure model. Different ablation pressure profiles, including Gaussian-shaped, cosine-shaped, and focused-on-point, were found to drive the droplet into different shapes. To determine the dependence of droplet deformation on the pressure pulse duration at constant total momentum, an analytical acoustic model (assuming small density fluctuations) was further derived for the internal pressure, pressure impulse, and velocity fields~\citep{reijers2017}. However, it was reported that tin sheets produced in experiments and most advanced EUV light sources often curve in a direction opposite to the theoretical predictions~\citep{kurilovich2016, hernandez2022, francca2025}. An instantaneous pressure impulse described by a raised cosine function was introduced to be able to reproduce the curvature inversion in the simulations~\citep{francca2025}. Droplet fragmentation was investigated via two-fluid simulations also in the framework of applying a pressure impulse on an incompressible droplet~\citep{nykteri2022}. The results showed a good agreement with the experimental observations of~\cite{klein2015} with respect to the expansion of the liquid sheet and the development of a polydisperse cloud of fragments. Although the pressure-impulse modelings are convenient to be employed in simulations on the droplet dynamics, the pressure impulse profiles (spatial and temporal) need to be empirically prescribed. Moreover, it is challenging to consider the dynamic feedback of the droplet to the plasma, which is expected to adjust the pressure profiles in flight. 

On the other hand, droplet dynamics can be obtained from direct numerical simulations that comprehensively take all the important physical processes into account. Radiation-hydrodynamic (RHD) codes such as FLASH~\citep{fryxell2000}, HEIGHTS~\citep{Sizyuk2015},  HELIOS-CR~\citep{macFarlane2006}, RALEF~\citep{basko2012}, and RHDLPP~\citep{min2024}, consider detailed modelings on the key LPP-relevant processes, including laser absorption, radiation transport, and heat conduction, and they have been used for LPP-EUV simulations~\citep{sheil2023}. \citet{kurilovich2018} performed simulations of interactions between a nanosecond pre-pulse and a tin droplet using two-dimensional (2D) RALEF code, and obtained a power-law scaling of propulsion velocity versus laser energy which agrees well with experimental data. The ratio of propulsion speed and initial radial expansion rate on a broad range of parameters (including laser energies, spot sizes, and droplet sizes) given by the RALEF simulations~\citep{hernandez2022} were shown to well agree with the experiments, and the energy partitioning between the deformation and the propulsion of the droplet was consistent with the instantaneous pressure-driven droplet simulations by~\cite{gelderblom2016}. 
Nevertheless, the RHD modelings were based on a single-fluid framework and required a unified equation of state (EOS) like the Frankfurt EOS (FEOS)~\citep{faik2012} to model tin coexisting in various states during the pre-pulse stage. While the RHD codes have been shown effective for simulating the laser-plasma interactions and the early phase of the droplet deformation, it was recently reported that a RHD code could not adequately simulate late-time droplet deformation~\citep{francca2025}. This shortcoming likely explains why investigations into the later droplet dynamics such as thin-sheet formation and fragmentation~\citep{klein2020, liu2022} still largely rely on pure hydrodynamics simulations~\citep{liu2017}. Therefore, the development of a radiation two-phase flow model, capable of simulating plasma physics, long-term evolution of droplet and their interactions, is a critical objective for LPP-EUV research.

In this paper, we propose a radiation two-phase flow model for the plasma-liquid interactions, based on a diffuse interface methodology. Specifically, we integrate radiation hydrodynamics for the plasma with the Euler equations for a compressible liquid, by extending a five-equation diffuse interface formulation~\citep{allaire2002, kapila2001} to incorporate key physics: radiation transport, thermal conduction, and ionization. In particular, the model enforces pressure and velocity equilibrium across the diffuse interface region, and can handle two different fluids with large density contrasts. To ensure physical fidelity, we develop closure models for energy fluxes and sources within the interface region. These closures guarantee the correct jump conditions at interfaces and ensure the model asymptotically recovers the pure-phase governing equations in bulk regions. We validate the model by demonstrating its accurate treatment of radiation transport and thermal conduction, and confirm that it correctly reduces to compressible two-phase flows in the absence of radiative effects. Finally, to evaluate its capability for self-consistently coupling laser-plasma physics with compressible droplet dynamics, we apply the radiation two-phase flow model to simulate a benchmark pre-pulse scenario, where a $50\ \upmu \mathrm{m}$ tin droplet is irradiated by a $10 \  \mathrm{ns}$ laser pulse.

The rest of this paper is organized as follows. Section 2 introduces the governing equations and equations of state for the bulk fluids. In Section 3, a new diffuse interface model is derived by unifying the governing equations for the liquid and plasma phases, and appropriate mixture closure relations are proposed. Section 4 presents a suitable numerical scheme for solving the model. Section 5 provides validation through several test cases: simulations of radiation-plasma flows and liquid-gas two-phase flows are compared against established results from the literature. Subsequently, the model is applied to simulate the deformation of a tin droplet irradiated by a nanosecond laser pulse. Finally, Section 6 summarizes the conclusions of this work.

\section{Governing equations and equation of state for bulk fluids}
\label{sec:background}
In this section, we present the governing equations and equation of state for the plasma and the liquid considered. In particular, we use the radiation-hydrodynamic model for the plasma and the Euler equations for the liquid.

\subsection{Radiation hydrodynamics for the plasma}
\label{subsec:rad-hydro}
A plasma in the presence of radiation photons can be modeled with a set of radiation hydrodynamics equations in the following form:
\begin{equation}
    \begin{cases}
    \partial_{t} \rho_{g} +\nabla \cdot \left ( \rho_{g}\mathbf{u}_{g} \right ) =0 , \\
    \partial_{t} \left ( \rho_{g}\mathbf{u}_{g} \right )  +\nabla \cdot \left ( \rho_{g} \mathbf{u}_{g}\mathbf{u}_{g} \right )+\nabla \left(p_{g}+p_{r}\right) =0 , \\
    \partial_{t} \left ( \rho_{g} e_{g} \right )  +\nabla \cdot \left ( \rho_{g} e_{g} \mathbf{u}_{g} \right ) +p_{g}\nabla \cdot \mathbf{u}_{g}=\nabla \cdot \left ( \kappa_{g}\nabla T_{g} \right )  +\omega_{r,g}\left ( T_{r}^{4}-T_{g}^{4} \right )  , \\
    \partial_{t} E_{r} +\nabla \cdot \left ( E_{r} \mathbf{u} \right ) +p_{r}\nabla \cdot \mathbf{u}=\nabla \cdot  \left ( \chi_{r,g}\nabla T_{r}^{4} \right ) + \omega_{r,g}\left ( T_{g}^{4}-T_{r}^{4}\right ),
    \end{cases}
    \label{eq:rad}
\end{equation}
where the subscripts $g$ and $r$ denote plasma (i.e. ionized gas) and radiation respectively hereafter. $t$ is the time, and the plasma quantities include $\rho_{g}$ the mass density, $\mathbf{u}_{g}$ the macroscopic velocity, $p_{g}$ the pressure, $T_{g}$ the temperature, $e_{g}$ the specific internal energy, and $\kappa_{g}$ the thermal conduction coefficient. The two components of the plasma, i.e. electrons and ions, are assumed to have identical macroscopic fluid velocity and temperature and act as a single fluid in this model. Radiation has been modeled under a diffusion approximation, i.e. optically thick~\citep{castor2004}. The radiation photons act as a special ideal gas without the stationary mass and provide a radiation pressure $p_r$. They are assumed to follow the Planckian (blackbody) distribution with a radiation temperature $T_r$ that is allowed to differ from the local plasma temperature $T_g$. The radiation energy $E_r = a T_{r}^{4}$ ($a=7.5657\times10^{-16}\ \mathrm{J}\cdot \mathrm{m}^{-3}\cdot \mathrm{K}^{-4}$ is the radiation constant) given by the Planckian distribution is linked with $p_r$ as $p_{r} = E_{r}/3$ due to the Eddington approximation~\citep{castor2004}. Once $T_r$ differs from $T_g$, the radiation photons exchange energy with the plasma at an energy-exchange rate of $\omega_{r,g}$. 
Under the diffusion approximation, part of the energy transport carried by radiation photons effectively acts like an energy diffusion with a coefficient $\chi_{r,g}$, while the other part acts as the energy convection together with the plasma, as demonstrated in the last equation of Eq.~(\ref{eq:rad}). 

The last two equations in Eq.~(\ref{eq:rad}) are the energy equations for the plasma and the radiation, respectively. Alternatively, the combination of these two equations yields the equation for the total energy of the system,
\begin{align}
    \partial _{t} \left [ \rho_{g}\left ( e_{g}+\left | \mathbf{u}_{g} \right |^{2}/2 \right )+E_{r}\right ] +&\nabla \cdot \left [ \rho_{g}\left ( e_{g}+\left | \mathbf{u}_{g} \right |^{2}/2 \right ) \mathbf{u}_g +E_{r} \mathbf{u}_{g}\right ] + \nabla \cdot\left [ \left ( p_{g}+p_{r} \right )\mathbf{u}_g \right ]  = \nonumber \\
    &\nabla \cdot \left ( \kappa_{g}\nabla T_{g} \right ) + \nabla \cdot \left ( \chi_{r,g}\nabla T_{r}^{4} \right ).
    \label{eq:rad_ene}
\end{align}

\subsection{Euler equations for the liquid}
\label{subsec:Euler}
The liquid is assumed to be inviscid and weakly compressible, and its dynamics is governed by the Euler equations:
\begin{equation}
    \begin{cases}
    \partial_t \rho_{l} + \nabla \cdot (\rho_{l} \mathbf{u}_{l}) = 0, \\
    \partial_t (\rho_{l} \mathbf{u}_{l}) + \nabla \cdot (\rho_{l} \mathbf{u}_{l}\mathbf{u}_{l}) + \nabla p_{l} = 0, \\
    \partial_t [\rho_{l} (e_{l}+\left | \mathbf{u}_{l} \right |^{2}/2) ] + \nabla \cdot [\rho_{l} (e_{l}+\left | \mathbf{u}_{l} \right |^{2}/2 ) \mathbf{u}_{l}] + \nabla \cdot (p_{l}\mathbf{u}_{l}) = 0,
    \end{cases}
    \label{eq:Euler}
\end{equation}
where subscript $l$ denotes liquid hereafter in this work. $\rho_{l}$, $\mathbf{u}_{l}$, $p_{l}$, and $e_{l}$ are density, velocity, pressure, and specific internal energy of the liquid, respectively. In order to be consistent with the radiation hydrodynamics equations, Eq.~(\ref{eq:rad}), the energy equation in Eq.~(\ref{eq:Euler}) is rewritten in a non-conservative form:
\begin{equation}
    \partial_t (\rho_{l} e_{l}) + \nabla \cdot (\rho_{l} e_{l} \mathbf{u}_{l}) + p_{l}\nabla \cdot \mathbf{u}_{l} = 0.
    \label{eq:liquid_ene}
\end{equation}

The liquid is assumed to be very opaque to radiation, and the self-emission of the liquid is negligibly weak due to its low temperature. Although placed in an radiative environment, the radiation energy of and the radiation transport inside the liquid are neglected, and thus are absent in Eq.~(\ref{eq:Euler}). So does heat conduction in the liquid, which is also negligible compared to that in the plasma.

\subsection{Equations of state}
\label{subsec:EOS}
The plasma and the liquid have different material properties and thus obey different EOS. We employ the stiffened gas (SG) model~\citep{shyue1998}, which is able to describe EOS of different materials in a unified form by tuning parameters, for both fluids.

The EOS of the liquid reads
\begin{equation}
    \begin{cases}
    p_{l}+\gamma_{l}\mathrm{P}_{\infty,l} = \left(\gamma_{l} -1\right)\rho_{l} e_{l},\\
    \rho_{l} e_{l} = \rho_{l} c_{v,l} T_{l}+\mathrm{P}_{\infty,l},
    \end{cases}
    \label{eq:SG_EOS}
\end{equation}
where the specific heat ratio $\gamma_{l}$, the reference pressure $\mathrm{P}_{\infty,l}$, and the specified heat capacity at constant volume $c_{v,l}$~\citep{lemetayer2004} are three constant parameters that need to be assigned. The values of $\gamma_{l}$ and $\mathrm{P}_{\infty,l}$ are usually chosen such that the sound speed inside the liquid
\begin{equation}
    C_{l}=\sqrt{\gamma_{l}\left(p_{l}+ \mathrm{P}_{\infty,l}\right)/\rho_{l}}
    \label{eq:SG_cl}
\end{equation}
is close to reality~\citep{shyue1998}. Typical choices of $\gamma_{l}$, $\mathrm{P}_{\infty,l}$ and $c_{v,l}$ for a couple of liquids in present work are listed in Table~\ref{tab:SGEOS_Para}, to recover a reference state under the standard atmospheric pressure: $T_l = 300$ K, $\rho_{l} = 1000\ \mathrm{kg/m^{3}}$, $C_{l} = 1450\ \mathrm{m/s}$ for water; $T_l = 593$ K, $\rho_{l} = 6980\ \mathrm{kg/m^{3}}$, $C_{l} = 2471\ \mathrm{m/s}$ for liquid-tin.
\begin{table}
    \centering
    \begin{tabular}{cccc}
    Fluid & \qquad $\gamma_{l}$ \qquad & \qquad $\mathrm{P}_{\infty,l}\left ( \mathrm{Pa} \right )$ \qquad & \qquad $c_{v,l}\left ( \mathrm{J}/\left(\mathrm{kg}\cdot\mathrm{K}\right) \right )$ \qquad  \\
    Water  & \qquad $4.4$ \qquad & \qquad $6\times10^{8}$ \qquad & \qquad $1816$ \qquad \\
    Liquid-Tin  & \qquad $30$ \qquad & \qquad $1.421\times10^{9}$ \qquad & \qquad $210$ \qquad \\
    \end{tabular}
    \caption{SG parameters for two liquids considered in the present work.}
    \label{tab:SGEOS_Para}
\end{table}

The plasma is modeled as an ideal gas, corresponding to a particular SG with the reference pressure $\mathrm{P}_{\infty,g}$ set to zero. Its EOS reads
\begin{equation}
    \begin{cases}
    p_{g} = \left(\gamma_{g} -1\right)\rho_{g} e_{g},\\
    \rho_{g} e_{g} = \rho_{g} c_{v,g} T_{g},
    \end{cases}
    \label{eq:Ideal_Gas}
\end{equation}
where the specific heat ratio $\gamma_{g}$ is set to $5/3$. $c_{v,g}$ is the specified capacity at constant volume considering the average ionization degree ($Z_{g}$) as
\begin{equation}
    c_{v,g} = \frac{R \left(1+Z_{g}\right) }{\left(\gamma_{g}-1\right)M_{g}}.
    \label{eq:Ideal_cv}
\end{equation}
Here $R=8.314\ \mathrm{J}/\left(\mathrm{mol}\cdot\mathrm{K}\right)$ is the universal gas constant and $M_{g}$ is the molar mass of the element ($118.71\ \mathrm{g}/\mathrm{mol}$ for tin). The sound speed in the plasma is written in a similar form as Eq.~(\ref{eq:SG_cl}):
\begin{equation}
    C_{g}=\sqrt{\gamma_{g} p_{g}/\rho_{g}}.
    \label{eq:SG_cg}
\end{equation}
Note that $C_g$ can be an order of magnitude larger than $C_l$, e.g. a typical tin plasma emitting the interested EUV lights has $T_g$ ranging from $10^{5}$ to $5\times10^{5}$ $\mathrm{K}$~\citep{nishihara2008}, yielding $C_g$ from $11000$ to $25000$ $\mathrm{m/s}$. 

\section{Diffuse interface model}
\label{sec:diffuse}
In this section, we establish a diffuse interface model for plasma-liquid interactions, and particularly focus on the exchange of momentum and energy between the plasma and the liquid. 

\subsection{Jump conditions at interfaces}
\label{subsec:condition}
In this work, we model the interface between plasma and liquid as a sharp contact discontinuity with no phase transition or chemical reaction, as illustrated in figure~\ref{fig:DiffuseInterface}($a$). The two phases are spatially separated and immiscible, with distinctly different material properties, thereby leading to jump conditions at interfaces. Since the interface is essentially a contact discontinuity, the Rankine-Hugoniot conditions should be strictly satisfied, which are the embodiment of the conservation laws at discontinuities. Specifically, the normal velocity and pressure are continuous across the interface: ${\bf u}_l \cdot {\bf n} ={\bf u}_g \cdot {\bf n}$ and $p_l=p_g$, where ${\bf n}$ is unit normal vector to the interface. Apart from the Rankine-Hugoniot conditions, the thermal fluxes in the plasma are assumed to vanish at interfaces, due to the comparatively negligible thermal conduction coefficient of the liquid. Similarly, the radiation energy fluxes are also expected to vanish at the interface because of high opacity of the liquid. In the present study, all these conditions are well handled by the proposed diffuse interface model introduced in $\S$~\ref{subsec:5equation} together with the mixing rules introduced in $\S$~\ref{subsec:closure} and $\S$~\ref{subsec:diffuseflux}.

\subsection{Diffuse interfaces}
\label{subsec:5equation}
\begin{figure}
\begin{center}
\includegraphics[width=13cm]{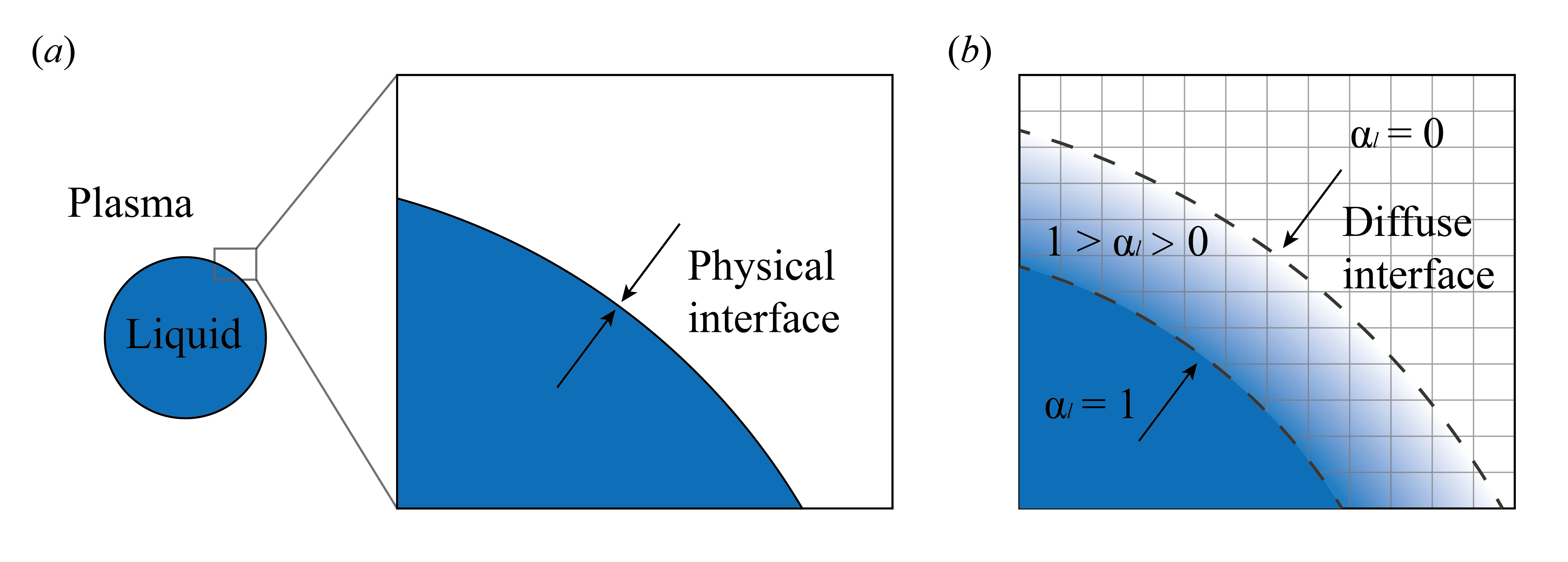}
\caption{($a$) Schematic of an interface separating the immiscible plasma and liquid, where the blue phase represents the liquid and the white phase represents the plasma; ($b$) A diffuse interface is used to replace the physical interface on a Cartesian grid, and the volume fraction of the liquid $\alpha_{l}$ is adopted to represent the interface position, where $0\le\alpha_{l}\le 1$.}
\label{fig:DiffuseInterface}
\end{center}
\end{figure}

In diffuse interface models, the sharp interface between two immiscible fluids is replaced by a diffuse interface with finite thickness where the two fluids are mixed, and the jump conditions at the interface is realized by mixing rules in the diffuse interface region. Inspired by the concept of diffuse interface, specifically the transport five-equation model~\citep{allaire2002}, we propose a diffuse interface model for liquid-plasma interactions, where the volume fraction of the liquid ($\alpha_l$) is used to represent the interface position, as illustrated in figure~\ref{fig:DiffuseInterface}($b$). We can see that $\alpha_l$ rapidly changes from 1 on the liquid side to 0 on the plasma side within a thin layer, which is referred hereafter to as the diffuse interface region. In the diffuse interface region the fluid is treated as a homogeneous mixture of the liquid and the plasma.

Similar to the model of~\cite{allaire2002}, the proposed model assumes that the velocities and pressures of the two phases are in equilibrium (namely mechanical equilibrium) in the diffuse interface region, i.e. the mixture velocity $\mathbf{u} = \mathbf{u}_l = \mathbf{u}_g$, and the mixture pressure $p = p_l = p_g$, but can be in thermal non-equilibrium ($T_l \ne T_g$). Such mechanical equilibrium at the interface ensures the proper propagation of acoustic waves across material interfaces while allowing the fluids to have different densities, internal energies, and equations of state. As a result, we can tentatively write the diffuse interface model for liquid-plasma interactions, which consists of an interface evolution equation, two phasic mass equations, a mixture momentum equation, a mixture energy equation and a radiation energy equation (Note that the liquid-phase region is also encompassed by the radiation field, with the radiation energy being set to zero),
\begin{equation}
    \begin{cases}
    \partial_{t} \alpha_{l}+\mathbf{u} \cdot \nabla \alpha_{l}= 0 , \\
    \partial_{t} \left ( \rho_{l} \alpha_{l} \right )  +\nabla \cdot \left ( \rho_{l} \alpha_{l}\mathbf{u}\right ) =0 ,\\
    \partial_{t} \left ( \rho_{g} \alpha_{g} \right )  +\nabla \cdot \left ( \rho_{g} \alpha_{g}\mathbf{u}\right ) =0 ,\\
    \partial_{t} \left ( \rho\mathbf{u} \right )  +\nabla \cdot \left ( \rho \mathbf{u}\mathbf{u} \right )+\nabla \left(p+p_{r}\right) =0, \\
    \partial_{t} \left ( \rho e \right )  +\nabla \cdot \left ( \rho e \mathbf{u} \right ) +p\nabla\cdot \mathbf{u}= -\nabla\cdot \mathbf{q}+ S_{r} , \\
    \partial_{t} E_{r}  +\nabla \cdot \left ( E_{r} \mathbf{u} \right ) +p_{r}\nabla\cdot \mathbf{u}=-\nabla\cdot \mathbf{F}_{r} -S_{r},
    \end{cases}
    \label{eq:Proposed}
\end{equation}
where $\rho$ and $e$ are the density and the specific internal energy of the mixture, respectively. The energy flux due to thermal conduction $\mathbf{q}$, the energy deposition rate to the mixture from radiation $S_{r}$, and the energy flux due to radiation diffusion $\mathbf{F}_{r}$, are applied to the mixture; their detailed expressions are provided in $\S$~\ref{subsec:diffuseflux}. 
It is worthy to note that the proposed model of Eq.~(\ref{eq:Proposed}) reduces to the original transport five-equation model for compressible two-phase flows~\citep{allaire2002} in the absence of thermal conduction and radiation transport (i.e. $p_{r}$, $E_{r}$, $S_{r}$, $\mathbf{q}$, and $\mathbf{F}_{r}$ are all zero). Eq.~(\ref{eq:Proposed}) are not closed yet until the EOS of the mixture is supplied.

\subsection{Equation of state in the diffuse interface region}
\label{subsec:closure}
In the absence of phase transition and chemical reaction, we establish the EOS for the mixture in a similar manner as~\cite{allaire2002}. In the diffuse interface region, the conservation of mass and energy straightforwardly leads to $\rho$ and $e$ in the form of
\begin{gather}
    \rho = \rho_{l}\alpha_{l}+\rho_{g}\alpha_{g}, 
    \label{eq:rhoalpha}\\
    \rho e = \rho_{l}\alpha_{l}e_{l}+\rho_{g}\alpha_{g}e_{g},
    \label{eq:rhoe}
\end{gather}
respectively. 
The isobaric closure~\citep{allaire2002} is adopted, namely assuming the mixture is in mechanical (pressure) equilibrium:
\begin{equation}
    p = p_{l}\left(\rho_{l}, e_{l}\right)=p_{g}\left(\rho_{g},   e_{g}\right).
    \label{eq:isobaric}
\end{equation}
Substituting the EOSs of the liquid (Eq.~(\ref{eq:SG_EOS})) and the plasma (Eq.~(\ref{eq:Ideal_Gas})) into Eq.~(\ref{eq:rhoe}) under the pressure equilibrium condition of Eq.~(\ref{eq:isobaric}) readily yields the EOS for the mixture:
\begin{equation}
    \left(\displaystyle \frac{\alpha_{l}}{\gamma_{l}-1} +\frac{\alpha_{g}}{\gamma_{g}-1}\right)p=\rho e-\left ( \displaystyle \frac{\alpha_{l}\gamma_{l}\mathrm{P}_{\infty,l}}{\gamma_{l}-1}+\frac{\alpha_{g}\gamma_{g}\mathrm{P}_{\infty,g}}{\gamma_{g}-1} \right ),
    \label{eq:Mix_EOS}
\end{equation}
which can also be expressed in a neater form similar to the stiffened gas model as 
\begin{equation}
    p+\gamma \mathrm{P}_{{\infty}}=\left(\gamma-1\right)\rho e,
    \label{eq:mixture_EOS}
\end{equation}
where the specific heat ratio $\gamma$ and the reference pressure $\mathrm{P}_{\infty}$ of the mixture are defined as:
\begin{equation}
    \left\{\begin{matrix}
    \displaystyle\frac{1}{\gamma - 1}  = \frac{\alpha_{l}}{\gamma_l - 1} + \frac{\alpha_{g}}{\gamma_g - 1}\ , 
    \\[10pt]
    \displaystyle\frac{\gamma \mathrm{P}_{\infty}}{\gamma - 1}  = \frac{\alpha_{l} \gamma_l \mathrm{P}_{\infty,l}}{\gamma_l - 1} + \frac{\alpha_{g} \gamma_g \mathrm{P}_{\infty,g}}{\gamma_g - 1}\ .
    \end{matrix}\right.
    \label{eq:GammaPinfty}
\end{equation}

The sound speed of the mixture ($C$), which will be needed by the numerical solver discussed later in $\S$ \ref{sec:numerical} as a characteristic speed, is constructed in a similar way as~\cite{allaire2002}:
\begin{equation}
    \frac{\rho C^{2}}{\gamma-1}=\frac{\alpha_{l}\rho_{l}C_{l}^2}{\gamma_{l}-1}+\frac{\alpha_{g}\rho_{g}C_{g}^2}{\gamma_{g}-1}.
\end{equation}
This formulation guarantees that the sound speed monotonically transits from one phase to the other across the diffuse interface region in present work.

\subsection{Fluxes and sources in the diffuse interface region}
\label{subsec:diffuseflux}
To complete Eq.~(\ref{eq:Proposed}), the energy fluxes and sources for the mixture (i.e. $\mathbf{q}$, $\mathbf{F}_r$, and $S_r$) need to be formulated, in addition to the EOS. Furthermore, their expressions must satisfy the jump conditions described in $\S$~\ref{subsec:condition}, i.e. these quantities vanish at the interfaces. 

To obtain physically reasonable expressions of $S_r$ and $\mathbf{q}$ in the mixture energy equation of Eq.~(\ref{eq:Proposed}), we start with the separate energy equation for each phase:
\begin{gather}
    \partial_{t} \left ( \rho_{l}\alpha_{l} e_{l} \right )  +\nabla \cdot \left ( \rho_{l}\alpha_{l} e_{l} \mathbf{u} \right ) +\alpha_{l}p\nabla\cdot \mathbf{u} = -\nabla\cdot(\mathbf{q}_{ll}+\mathbf{q}_{lg})+S_{r,l}+Q_{lg} ,
    \label{eq:Energy1}
    \\
    \partial_{t} \left ( \rho_{g}\alpha_{g} e_{g} \right )  +\nabla \cdot \left ( \rho_{g}\alpha_{g} e_{g} \mathbf{u} \right ) +\alpha_{g}p\nabla\cdot\mathbf{u} = -\nabla\cdot(\mathbf{q}_{gl}+\mathbf{q}_{gg})+S_{r,g}+Q_{gl} ,
    \label{eq:Energy2}
\end{gather}
where the fluxes and source for liquid are retained at this moment without losing generality; In other words, the liquid is also treated as a kind of plasma. Note that energy exchanges occur both within and between phases. Here, $\mathbf{q}_{ij}\ (i,\ j \in \{l,\ g\})$ denotes the thermal energy flux carried by the surrounding phase $j$ and applied to the internal phase $i$, $S_{r,i}\ (i \in \{l,\ g\})$ is the radiation energy deposition rate to phase $i$, and $Q_{ij} (i,\ j \in \{l,\ g\},\ i\ne j)$ is the energy exchange rate from internal phase $j$ to internal phase $i$ with $Q_{lg} = -Q_{gl}$. For generality, we initially retain all terms for the liquid phase, effectively treating it as a plasma with its own transport coefficients: the thermal conductivity $\kappa_{l}$, the radiative energy exchange rate $\omega_{r,l}$, and the radiative energy diffusion coefficient $\chi_{r,l}$, despite that these are typically much smaller than their plasma-phase counterparts.

The radiative energy source for each phase is modeled as
\begin{equation}
    S_{r,i}=\alpha_{i}\omega_{r,i}\left ( T_{r}^{4}-T_{i}^{4} \right ).  \quad i \in \{l,\ g\}
\end{equation}
Therefore the source of the mixture is the sum of $S_{r,g}$ and $S_{r,l}$ which reads:
\begin{equation}
    S_{r}=\alpha_{g}\omega_{r,g}\left ( T_{r}^{4}-T_{g}^{4} \right )+\alpha_{l}\omega_{r,l}\left ( T_{r}^{4}-T_{l}^{4} \right ).
    \label{eq:ene_source}
\end{equation}

According to its definition, the plasma-to-plasma conductive heat flux $\mathbf{q}_{gg}$ should be proportional to the local area over which the internal plasma contacts the surrounding plasma. For a homogeneous mixture of plasma and liquid, the local fraction of contact area between the plasmas is equal to the product of the volume fractions on both sides, namely $\alpha_{g}^+ \alpha_{g}^-$, where the superscripts $+$ and $-$ denote the internal and external side of the contact surface, respectively. Given homogeneity, $\alpha_{g}=\alpha_{g}^+= \alpha_{g}^-$ and $\alpha_{l}=\alpha_{l}^+= \alpha_{l}^-$. Therefore, we can have
\begin{equation}
    \mathbf{q}_{gg} = -\alpha_{g}^2\kappa_{g}\nabla T_{g}.
\end{equation}

Similarly, the contact area fraction between the internal plasma and the surrounding liquid can be expressed as $\alpha_{g}^+ \alpha_{l}^-$. The interphase thermal conduction is expected to be determined by the two phases, and thus we phenomenologically assign the effective interphase thermal conduction coefficient as $\sqrt{\kappa_{g}\kappa_{l}}$, a geometric average of $\kappa_{g}$ and $\kappa_{l}$. Assuming a local temperature difference $T_{g}^{+}-T_{l}^{-}$ over a small length scale $\delta \mathbf{x}$, the liquid-to-plasma flux $\mathbf{q}_{gl}$ can be estimated by
\begin{equation}
    \mathbf{q}_{gl} = -\alpha_{g}^+\alpha_{l}^-\sqrt{\kappa_{g}\kappa_{l}} \left ( T_{g}^{+}-T_{l}^{-} \right ) /\delta \mathbf{x}.
\label{q_express}
\end{equation}
Eq.~(\ref{q_express}) can be reformulated in two equivalent ways: 
\begin{align*}
    \mathbf{q}_{gl} &= -\alpha_{g}^+\alpha_{l}^-\sqrt{\kappa_{g}\kappa_{l}} \left (\nabla T_{g}+(T_{g}^{-}-T_{l}^{-}) /\delta \mathbf{x} \right ), \qquad \texttt{or}\\
    &= -\alpha_{g}^+\alpha_{l}^-\sqrt{\kappa_{g}\kappa_{l}} \left ( \nabla T_{l}+(T_{g}^{+}-T_{l}^{+}) /\delta \mathbf{x}  \right ).
\end{align*}
In these forms, the local temperature difference between the plasma and liquid phases ($T_{g}^{+}-T_{l}^{+}$ or $T_{g}^{-}-T_{l}^{-}$) implies that $\mathbf{q}_{gl}$ is also related to the exchange rate of internal energy from plasma to liquid, $Q_{gl}$, evaluated on the internal and external side of the contact surface, respectively.

To keep with the notational symmetry, the liquid-to-liquid and plasma-to-liquid energy fluxes due to thermal conduction can be expressed as
\begin{equation}
    \mathbf{q}_{ll} = -\alpha_{l}^{2}\kappa_{l}\nabla T_{l}, \qquad
    \mathbf{q}_{lg} = -\alpha_{l}^+\alpha_{g}^-\sqrt{\kappa_{g}\kappa_{l}}\left ( T_{l}^{+}-T_{g}^{-} \right ) /\delta \mathbf{x}.
\end{equation}
The mixture thermal flux is then contributed by all of the fluxes within and between phases, i.e. $\mathbf{q}=\mathbf{q}_{gg}+\mathbf{q}_{gl}+\mathbf{q}_{lg}+\mathbf{q}_{ll}$. It is worth noting that the temperature differences between the plasma and liquid phases in $\mathbf{q}_{lg}$ and $\mathbf{q}_{gl}$ cancel upon summation. Consequently, the final expression for $\mathbf{q}$ can be simplified accordingly,
\begin{equation}
    \mathbf{q} = -\alpha_{g}^{2}\kappa_{g}\nabla T_{g}-\alpha_{g}\alpha_{l}\sqrt{\kappa_{g}\kappa_{l}}(\nabla T_{g}+\nabla T_{l})-\alpha_{l}^{2}\kappa_{l}\nabla T_{l},
    \label{eq:flux_heat}
\end{equation}
This modeling approach has the desirable property that it reduces to the Fick's law $\mathbf{q}=- \kappa \nabla T$ in the limit where the two phases are identical, thereby ensuring its consistency with the single-phase theory.

The energy flux due to radiation diffusion is analogous to the conductive heat flux to some degree. In the optically thick limit, where the radiation can be described as a diffusion field, the radiation diffusion fluxes are proportional to the contact area between phases and the respective radiation diffusion coefficients ($\chi_{r,g},\ \chi_{r,l}$). By direct analogy to the conductive heat flux form derived in Eq.~(\ref{eq:flux_heat}), the radiation energy flux for the mixture can be written as
\begin{equation}
    \mathbf{F}_{r} = -(\alpha_{g}^{2}\chi_{r,g}+2\alpha_{g}\alpha_{l}\sqrt{\chi_{r,g}\chi_{r,l}}+\alpha_{l}^{2}\chi_{r,l})\,\nabla T_{r}^{4}.
    \label{eq:flux_rad}
\end{equation}

Because of the comparatively negligible thermal conduction coefficient and high opacity of the liquid, we set $\kappa_{l}\approx 0$, $\omega_{r,l}\approx 0$, and $\chi_{r,l}\approx 0$ in this paper. Accordingly, the radiative energy source, the conductive heat flux and the radiative diffusion flux yields a simplified model valid within the diffuse interface region
\begin{equation}
    \begin{cases}
    \mathbf{q}=-\alpha_{g}^{2}\kappa_{g} \nabla T_{g},\\
    S_{r}=\alpha_{g}\omega_{r,g}\left ( T_{r}^{4}-T_{g}^{4} \right ), \\
    \mathbf{F}_{r}=-\alpha_{g}^{2}\chi_{r,g} \nabla T_{r}^{4}.
    \end{cases}
    \label{eq:flux_source}
\end{equation}
This simplified formulation preserves a continuous transition for the energy fluxes and sources from the plasma to the liquid, governed by the plasma volume fraction $\alpha_{g}$. At the same time, it satisfies the required jump conditions at the interfaces.

\subsection{Summary of the model for plasma-liquid interactions} 
\label{subsec:Summary}
By applying the expressions of the energy fluxes and source in Eq.~(\ref{eq:flux_source}), we now rewrite the radiation two-phase flow model in Eq.~(\ref{eq:Proposed}) as
\begin{equation}
    \begin{cases}
    \partial_{t} \alpha_{l}+\mathbf{u} \cdot \nabla \alpha_{l}= 0 , \\
    \partial_{t} \left ( \rho_{l} \alpha_{l} \right )  +\nabla \cdot \left ( \rho_{l} \alpha_{l}\mathbf{u}\right ) =0 ,\\
    \partial_{t} \left ( \rho_{g} \alpha_{g} \right )  +\nabla \cdot \left ( \rho_{g} \alpha_{g}\mathbf{u}\right ) =0 ,\\
    \partial_{t} \left ( \rho\mathbf{u} \right )  +\nabla \cdot \left ( \rho \mathbf{u}\mathbf{u} \right )+\nabla \cdot\left((p+p_{r}) {\bf I}\right) =0, \\
    \partial_{t} \left ( \rho e \right )  +\nabla \cdot \left ( \rho e \mathbf{u} \right ) +p\nabla\cdot \mathbf{u}= \nabla\cdot (\alpha_{g}^{2}\kappa_{g} \nabla T_{g})+ \alpha_{g}\omega_{r,g}\left ( T_{r}^{4}-T_{g}^{4} \right ) , \\
    \partial_{t} E_{r}  +\nabla \cdot \left ( E_{r} \mathbf{u} \right ) +p_{r}\nabla\cdot \mathbf{u}=\nabla\cdot (\alpha_{g}^{2}\chi_{r,g} \nabla T_{r}^{4}) -\alpha_{g}\omega_{r,g}\left ( T_{r}^{4}-T_{g}^{4} \right ),
    \end{cases}
    \label{eq:summary}
\end{equation}
Although this model is derived for the mixture, it exhibits the correct asymptotic behavior by reducing to standard single-phase models in the limit of $\alpha_{g}$. It is straightforward that the system of equations in Eq.~(\ref{eq:summary}) converges to the Euler equation~(\ref{eq:Euler}) for the pure liquid in the limit of vanishing plasma and radiation, while it recovers the radiation hydrodynamics equations for a pure plasma (Eq.~\ref{eq:rad}) as $\alpha_{g} \rightarrow 1$.

\section{Numerical method}
\label{sec:numerical}
In order to numerically solve the radiation two-phase flow for plasma-liquid interactions described in Eq.~(\ref{eq:summary}), we employ an operator-splitting numerical algorithm~\citep{mcLachlan2002}. This approach is particularly effective for handling the multiphysics nature of the equations, where the characteristic timescales and numerical stiffness associated with convective transport differ significantly from those of diffusive processes. The method proceeds by separating the full system into two sequential sub-problems: A hyperbolic step and a parabolic step, and consequently allows us to use efficient numerical techniques to account for the distinct mathematical properties of each sub-problem. In the hyperbolic step, we integrate the subsystem governing convection of the fluid and radiation, Eq.~(\ref{eq:Convection}). 
\begin{subequations}
    \begin{align}
        &\partial_{t} \alpha_{l}+\nabla \cdot \left ( \alpha_{l}\mathbf{u}\right )= \alpha_{l} \nabla \cdot\mathbf{u} , \label{eq:Conv_part1} \\
        &\partial_{t} \left ( \rho_{l} \alpha_{l} \right )  +\nabla \cdot \left ( \rho_{l} \alpha_{l}\mathbf{u}\right ) =0 , \label{eq:Conv_part2} \\
        &\partial_{t} \left ( \rho_{g} \alpha_{g} \right )  +\nabla \cdot \left ( \rho_{g} \alpha_{g}\mathbf{u}\right ) =0 ,\label{eq:Conv_part3}\\
        &\partial_{t} \left ( \rho\mathbf{u} \right )  +\nabla \cdot \left ( \rho \mathbf{u}\mathbf{u} \right )+\nabla \cdot \left(\left(p+p_{r}\right) \mathbf{I}\right) =0, \label{eq:Conv_part4}\\
        &\partial_{t} \left ( \rho e \right )  +\nabla \cdot \left ( \rho e \mathbf{u} \right )+p\nabla \cdot \mathbf{u}=0 ,\label{eq:Conv_part5}\\
        &\partial_{t} E_{r}  +\nabla \cdot \left ( E_{r} \mathbf{u} \right ) +p_{r}\nabla \cdot \mathbf{u}=0. \label{eq:Conv_part6}
    \end{align}
    \label{eq:Convection}
\end{subequations}
\unskip\noindent
In the parabolic step, we integrate the subsystem governing thermal conduction and radiation diffusion, Eq.~(\ref{eq:Diffusion}).
\begin{equation}
    \begin{cases}
    \partial_{t} \left ( \rho e \right )  = \nabla \cdot \left ( \alpha_{g}^{2}\kappa_{g}\nabla T_{g} \right )+\alpha_{g}\omega_{r,g}\left ( T_{r}^{4}-T_{g}^{4} \right ), \\[6pt]
    \partial_{t} E_{r}   =\nabla \cdot \left ( \alpha_{g}^{2}\chi_{r,g}\nabla T_{r}^{4} \right ) + \alpha_{g}\omega_{r,g}\left ( T_{g}^{4}-T_{r}^{4} \right ) .
    \end{cases}
    \label{eq:Diffusion}
\end{equation}
In particular, an explicit scheme is implemented for the convection of fluid and radiation, while an implicit scheme for the thermal conduction and radiation diffusion; details are provided in the following subsections.

\subsection{Hyperbolic step: convection of fluid and radiation}
Note that the energy equations for fluid and radiation convection, i.e. Eqs.~(\ref{eq:Conv_part5}) and (\ref{eq:Conv_part6}), are not in the conservative form, primarily due to the pressure-work terms associated with fluid compressibility. In order to impose strict energy conservation, we instead consider the convection of the total energy density $\rho e_{t}$, which is defined as $\rho e_{t}=\rho e+E_{r}+\rho\left|\mathbf{u} \right|^{2}/2$. The governing equation for $\rho e_{t}$ is 
\begin{equation}
    \partial _{t} \left ( \rho e_{t}\right ) +\nabla \cdot \left ( \rho e_{t} \mathbf{u} \right ) + \nabla \cdot\left (  p_{t}\mathbf{u} \right) = 0 ,
    \label{eq:totale}
\end{equation} 
with the total pressure given by $p_{t}=p+p_{r}$. Eq.~(\ref{eq:totale}) is obtained by summing Eq.~(\ref{eq:Conv_part5}), Eq.~(\ref{eq:Conv_part6}), and the dot product of Eq.~(\ref{eq:Conv_part4}) with the velocity vector; further details are provided in Appendix~\ref{A}.

A new hyperbolic subsystem is thus formulated by Eq.~(\ref{eq:Convection}$a$-$d$) and Eq.~(\ref{eq:totale}), and is discretized using a second-order finite-volume scheme with the numerical fluxes computed via the HLLC approximate Riemann solver~\citep{toro2013}. The characteristic wave speed $C_{s}$ is obtained from the eigenvalue of the full system Eq.~(\ref{eq:Convection}), computed from its Jacobian matrix; see Appendix~\ref{B} for details. The term of velocity divergence in the volume-fraction advection equation (Eq.~\ref{eq:Conv_part1}) is evaluated using the adapt-HLLC scheme~\citep{johnsen2006}. The time step $\Delta t$ is chosen adaptively such that the Courant-Friedrichs-Lewy (CFL) condition is strictly satisfied,
\begin{equation}
    \Delta t=\mathrm{CFL}\cdot \min_{i\in D} \left \{ \frac{\sqrt[3]{V_{i}} }{\left | \mathbf{u}_{i} \right | +C_{s,i}}  \right \} ,
\end{equation}
where the $\mathrm{CFL}$ number ($=0.44$) is a prescribed constant, the subscripts $i$ denotes a Cartesian cell in the computational domain $D$, $C_{s}=\sqrt{C^{2}+4p_{r}/(3\rho)}$ is the local wave speed and $V$ is the volume of the cell.

After solving the fluid and radiation convection explicitly, we update the conservative variables to an intermediate state, denoted by the superscript $*$: [$\alpha_{l}^*$, $(\rho_{l}\alpha_{l})^*$, $(\rho_{g}\alpha_{g})^*$, $(\rho \mathbf{u})^*$, $(\rho e_{t})^*]^T$. The corresponding primitive variables at the intermediate state are also updated. From $(\rho e_{t})^*$, the internal energy $(\rho e)^*$ and radiation energy $(E_{r})^*$ of the mixture can then be calculated; see details in Appendix~\ref{C}.

\subsection{Parabolic step: thermal conduction and radiation diffusion}
An implicit scheme is adopted for the time integration of the thermal conduction and radiation diffusion in Eq.~(\ref{eq:Diffusion}). All the coefficients in the equations are linearized using the flow variables at the intermediate state:
\begin{equation}
    \begin{cases}
    \displaystyle\frac{(\rho e)^{n+1}-(\rho e)^{*}}{\Delta t}  = \nabla \cdot ((\alpha_{g}^{2}\kappa_{g})^{*}\,\nabla (T_{g})^{n+1})+(\alpha_{g}\omega_{r,g})^{*} \,( T_{r}^{4}-T_{g}^{4} )^{n+1},\\[10pt]
    \displaystyle\frac{ (E_{r} )^{n+1}- (E_{r} )^{*}}{\Delta t} =\nabla \cdot   (  (\alpha_{g}^{2}\chi_{r,g} )^{*}\,\nabla  (T_{r}^{4} )^{n+1}) +  (\alpha_{g}\omega_{r,g})^{*}\, (T_{g}^{4}-T_{r}^{4})^{n+1}.
    \end{cases}
    \label{eq:Diffusion_Num}
\end{equation}
The gradient and divergence terms are discretized using a central finite-difference scheme. By solving Eq.~(\ref{eq:Diffusion_Num}) iteratively, $\rho e$ and $E_{r}$ are updated from the intermediate state to the $n+1$ time step. Subsequently, by employing the equation of state, all remaining conservative and primitive variables associated with energy such as $p$, $p_{r}$, $T_{g}$, and $T_{r}$ are calculated for the $n+1$ time step, while those related to mass and momentum are retained from the intermediate state. This completes the numerical solution of the radiation two-phase flow system for plasma–liquid interactions.

\section{Model validation and numerical examples}
\label{sec:results}
In this section, we first validate the radiation two-phase flow model by simulating two types of flows: plasma single-phase flow and liquid-gas two-phase flow. The numerical results for each are compared against benchmark solutions from the literature. We then apply the proposed model to simulate a LPP-EUV relevant scenario, in which a nanosecond laser-pulse irradiates a tin droplet, leading to its deformation into a thin sheet.

\subsection{Radiative shock tube}
\label{subsec:vali-radiation}
\begin{figure}
\begin{center}
\includegraphics[width=12.5cm]{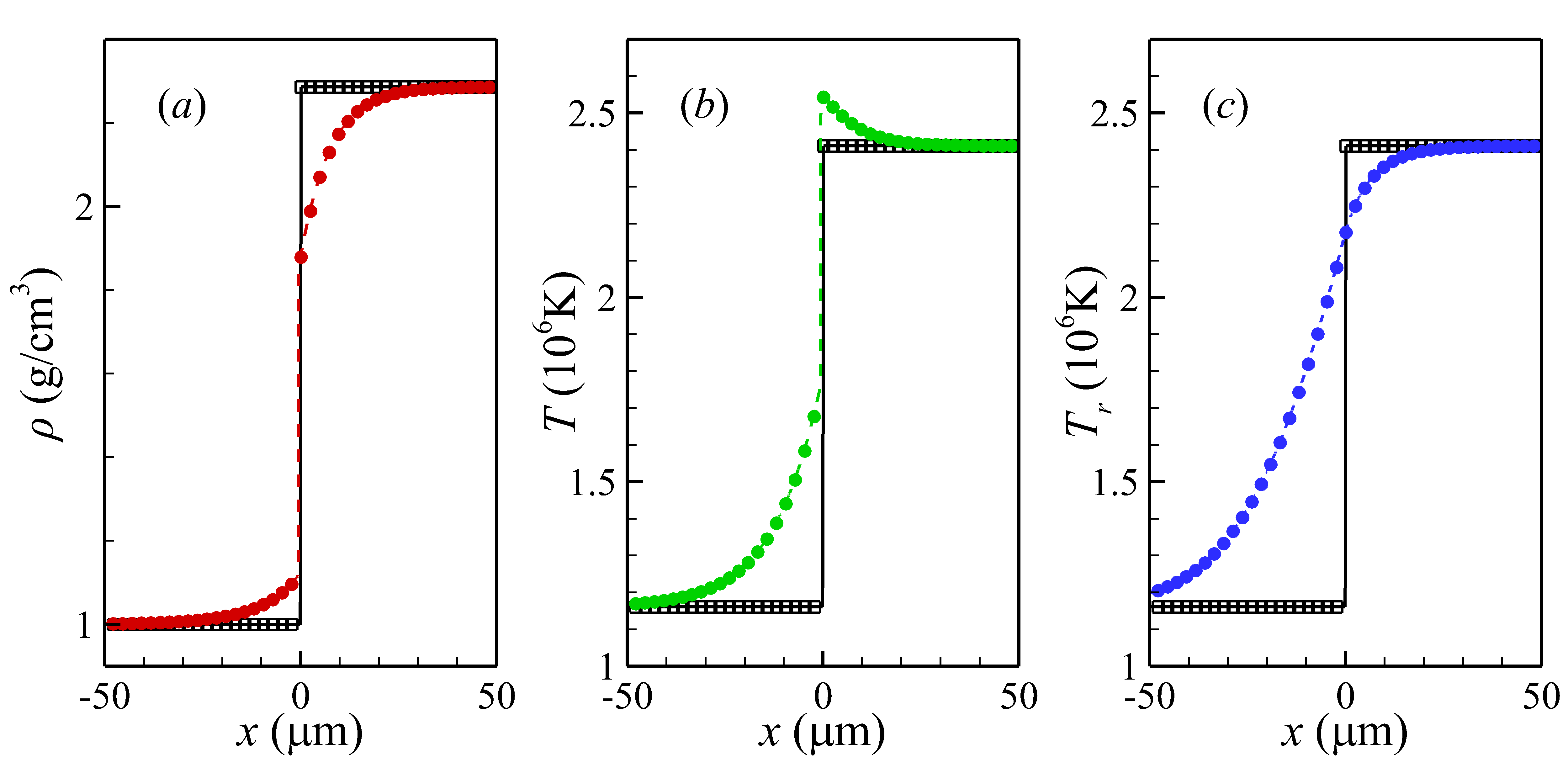}
\caption{The profiles of a radiative shock problem at $0\ \mathrm{ns}$ (black lines and open squares) and $4.2\ \mathrm{ns}$ (dashed lines and solid circles), in terms of density ($a$), plasma temperature ($b$), and radiation temperature ($c$). The symbols and lines denote numerical and semi-analytical solutions, respectively. Note that the symbols are sampled at every $8$th grid point for visual clarity.}
\label{fig:Test_Rad}
\end{center}
\end{figure}

To validate the radiation transport in our model, we consider a one-dimensional radiative shock tube problem, where the radiation energy fluxes and radiation pressure significantly affect the hydrodynamics. As illustrated in figure~\ref{fig:Test_Rad}, a low-temperature low-density plasma (with density $\rho^{L}$, temperature $T^{L}$ and velocity $u^{L}$) is initially impacting a high-temperature high-density plasma (with density $\rho^{R}$, temperature $T^{R}$ and velocity $u^{R}$). This configuration is designed to produce a stationary radiative shock at $x=0\ \upmu \mathrm{m}$, the solution of which can be compared with the semi-analytical result from~\cite{lowrie2008}. 

Because the fluid system only has one phase, i.e. plasma, we set $\alpha_{g}=1$ everywhere in the domain. The properties of the plasma are: an adiabatic index of $\gamma_{g} = 5/3$, a molar mass of $M_{g} = 2\ \mathrm{g/mol}$, and a constant ionization degree of $Z_{g} = 1$, which determine $c_{v,g} = 1.247\times10^8\ \mathrm{cm}^{2}/(\mathrm{s}^{2}\cdot\mathrm{K})$. Thermal conduction is neglected (i.e. $\kappa_{g}=0$). The diffusion coefficient and the exchange rate of the radiative energy are $\chi_{r,g} = 1.268\times10^{7}\ \mathrm{cm}^{2}/\mathrm{s}$ and $\omega_{r,g} = 1.268\times10^{13}\ \mathrm{s}^{-1}$, respectively. The initial conditions are defined by a step function at $x=0\ \upmu \mathrm{m}$, as illustrated in figure~\ref{fig:Test_Rad}. The radiation is initially set in thermal equilibrium with the plasma (i.e. $T_{r}=T$). The initial flow parameters of the plasma are listed as follows:
\begin{equation}
    \begin{cases}
    \rho^{R} = 2.286\ \mathrm{g}/\mathrm{cm}^{3},\ T^{R} = 2.411\times10^6\ \mathrm{K},\ u^{R} = 1.11\times10^7\ \mathrm{cm}/\mathrm{s}; \\
    \rho^{L} = 1\ \mathrm{g}/\mathrm{cm}^{3},\ T^{L} = 1.16\times10^6\ \mathrm{K},\ u^{L} = 2.536\times10^7\ \mathrm{cm}/\mathrm{s}.
    \end{cases}
\end{equation}
The computation is performed in a domain of $\left[-300,\ 300\right]\ \upmu \mathrm{m}$, discretized by a uniform grid of $2000$ points. 

Figure~\ref{fig:Test_Rad} presents numerical results at $t = 4.2\ \mathrm{ns}$ in terms of density, plasma temperature, and radiation temperature. These results are superimposed with the semi-analytical solution by~\cite{lowrie2008}. Obviously, the numerical results are in excellent agreement with the theoretical prediction, with respect to the position of the stationary shock and the diffusion profiles on either side of the discontinuity. The comparison successfully demonstrates that the proposed radiation two-phase model accurately captures complex plasma flow structures, including radiative shocks.

\subsection{Blast wave with thermal conduction}
\label{subsec:vali-heat}
\begin{figure}
\begin{center}
\includegraphics[width=12cm]{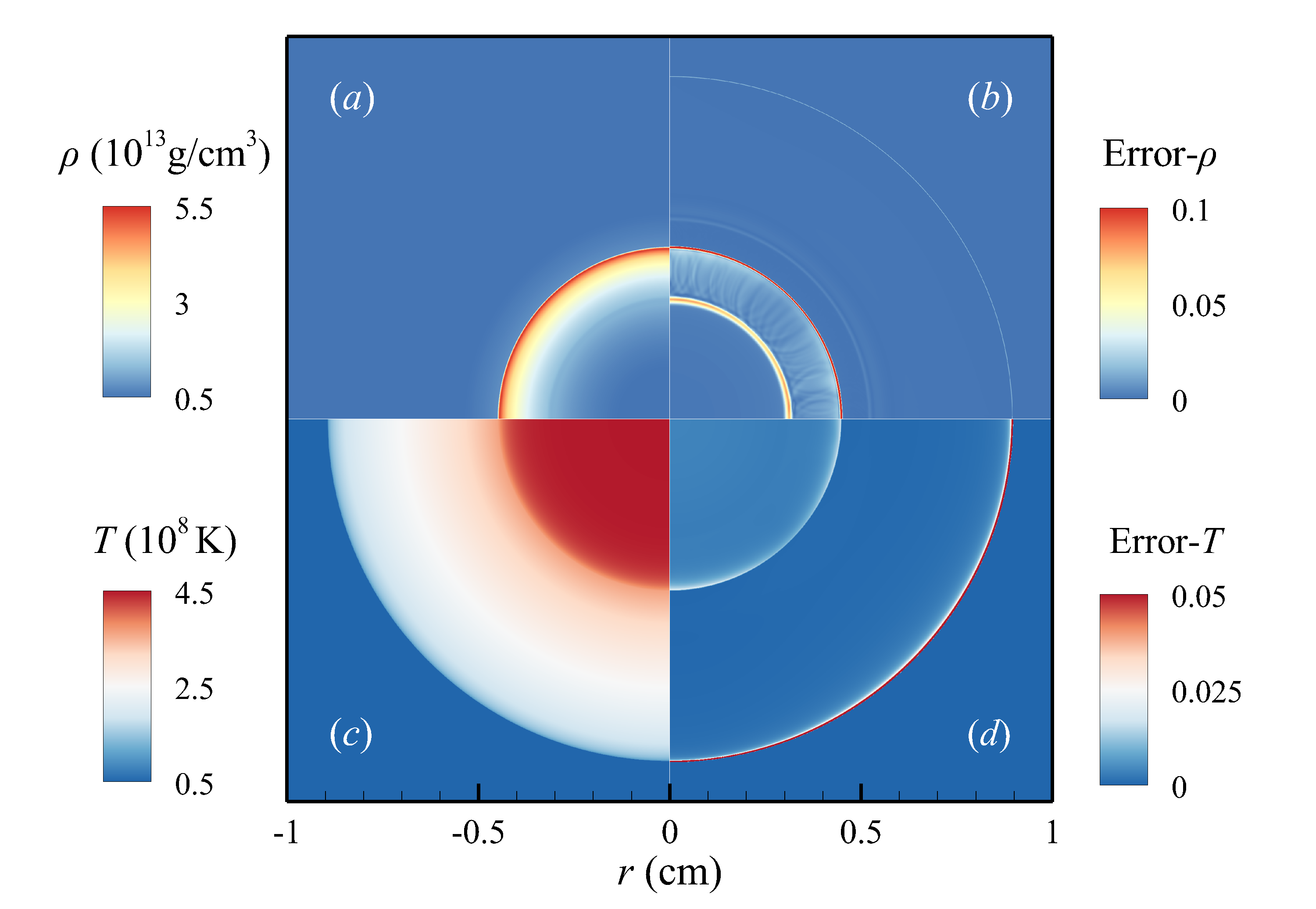}
\caption{Density ($a$), relative density error $\left|\rho^{N}-\rho^{A}\right|/\rho^{A}$ ($b$), temperature ($c$), and relative temperature error $\left|T^{N}-T^{A}\right|/T^{A}$ ($d$) for the Reinicke $\&$ Meyer-ter-vehn blast wave problem at $0.52\ \mathrm{ns}$, where superscripts $N$ and $A$ represent numerical and semi-analytical results, respectively.}
\label{fig:Test_Heat}
\end{center}
\end{figure}

To validate the thermal conduction in our model, we consider the Sedov-Taylor point explosion with nonlinear thermal conduction, also referred to as the Reinicke $\&$ Meyer-ter-vehn problem, which only involves the plasma phase and has the semi-analytical solution for the blast wave~\citep{reinicke1991}. To avoid the singularity of a point explosion, the initial condition for the simulation is taken from the semi-analytical solution right after the bang time. At this moment, the stationary and uniform ambient plasma is about to be swept through by a tiny and rapidly-expanding spherical blast wave. 

The plasma is modeled as an ideal gas with an adiabatic index of $\gamma_{g} = 1.25$, a molar mass of $M_{g} = 1\ \mathrm{g/mol}$ and an ionization degree of zero, giving a specific heat capacity at constant volume $c_{v,g} = 3.326\times10^8\ \mathrm{cm}^{2}/(\mathrm{s}^{2}\cdot\mathrm{K})$. The radiation transport is not excluded from this problem; accordingly, the radiation energy $E_{r}$, the radiation pressure $p_{r}$, the radiative energy exchange rate $\omega_{r,g}$, and the radiative energy diffusion coefficient $\chi_{r,g}$ for the plasma are all set to zero. The thermal conduction coefficient follows a nonlinear power-law function of density and temperature as recommended by~\cite{reinicke1991}, i.e. $\kappa_{g} = \rho_g^{-2}T_g^{6.5}\ (\mathrm{g}\cdot\mathrm{cm}\cdot\mathrm{s}^{-3}\cdot\mathrm{K}^{-1})$, where $\rho_{g}$ is in $\mathrm{g}/\mathrm{cm}^{3}$ and $T_{g}$ in $\mathrm{K}$. Axisymmetric simulations are performed in 2D cylindrical coordinates ($r,z$ plane). The initial condition corresponds to the semi-analytical solution at $t = 0.2\ \mathrm{ns}$, when the shock front has a radius of $0.225\ \mathrm{cm}$ and the heat front a radius of $0.45\ \mathrm{cm}$. The computational domain is $\left [ 0,\ 1 \right ] \times \left [ 0,\ 1 \right ]\ \mathrm{cm}$, and is discretized by $1024\times1024$ grid points. 

\label{subsec:vali-diffusion}
\begin{figure}
\begin{center}
\includegraphics[width=12cm]{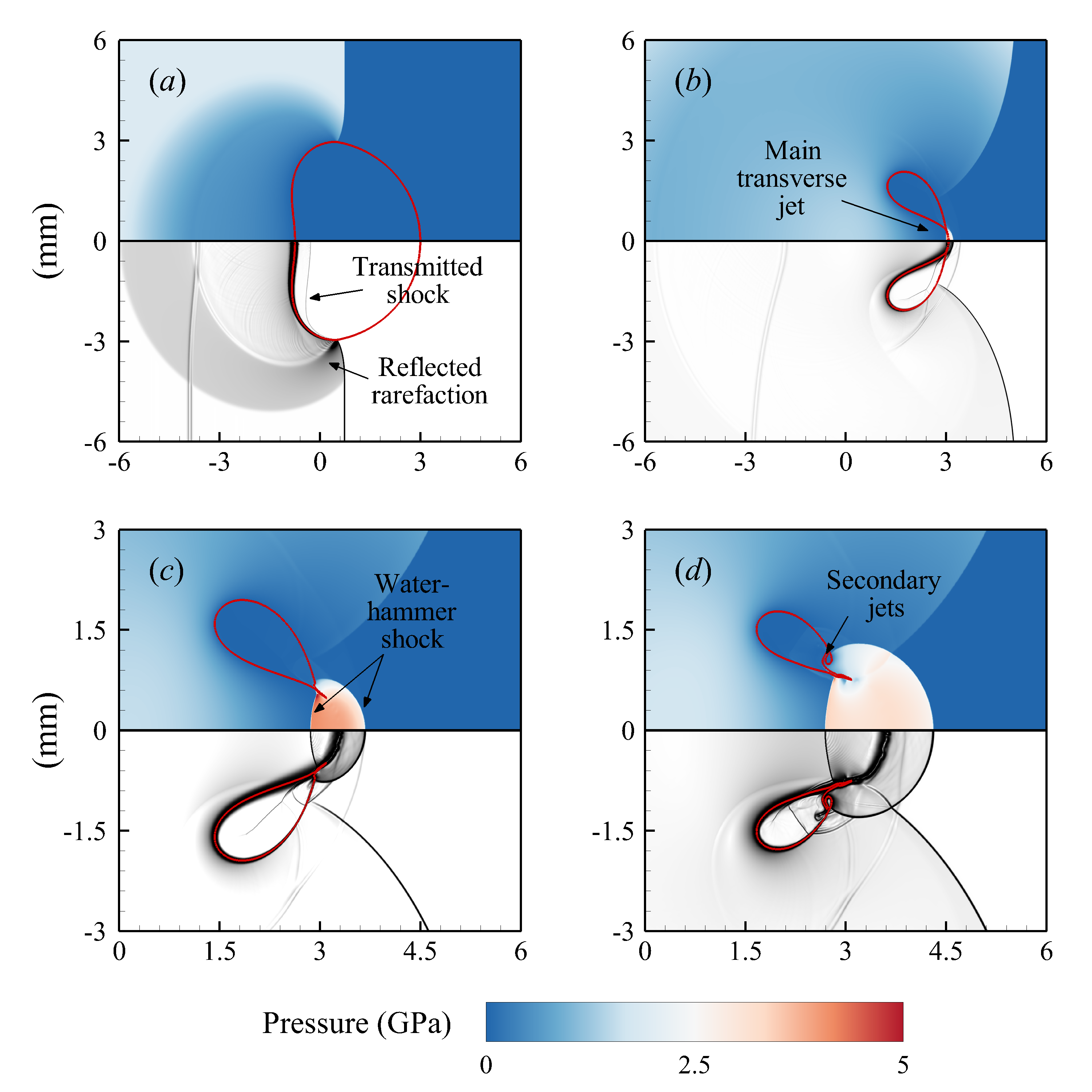}
%\put(-340, 330){\large \textbf{(a)}}
%\put(-10, 330){\large \textbf{(b)}}
\caption{Numerical results of bubble collapse induced by a planar shock with respect to pressure (upper half) and numerical schlieren (lower half) at $2.2\ \upmu \mathrm{s}$ ($a$), $3.7\ \upmu \mathrm{s}$ ($b$), $3.9\ \upmu \mathrm{s}$ ($c$), and $4.1\ \upmu \mathrm{s}$ ($d$), respectively. The red lines represent the bubble interfaces (by $\alpha_{g} = 0.5$).}
\label{fig:ShockBubble}
\end{center}
\end{figure}

Figure~\ref{fig:Test_Heat}($a$) and \ref{fig:Test_Heat}($c$) illustrate the distributions of density and temperature at $t = 0.52\ \mathrm{ns}$. As predicted by~\cite{reinicke1991}, the shock front reached a radius of $0.45\ \mathrm{cm}$ and the heat front $0.9\ \mathrm{cm}$. The corresponding relative errors in density and temperature are shown in figure~\ref{fig:Test_Heat}($b$) and \ref{fig:Test_Heat}($d$), respectively. The maximum errors, which do not exceed $10\%$, are localized at the discontinuities of the shock and heat front and are attributed to finite grid resolution. The good agreement with the semi-analytical solution demonstrates the capability of the proposed model to accurately deal with nonlinear thermal conduction.

\subsection{Bubble collapse induced by shockwave}
The shock-induced collapse of a gas bubble in water, where a planar shock wave propagating through the water interacts an underwater gas bubble, has been extensively studied by simulations~\citep{nourgaliev2006,hawker2012,bo2014}. This classic problem is used here to verify that the governing equations of the proposed model correctly reduce to those for compressible two-phase flows. In this case, thermal conduction and radiation transport are neglected. Consequently, all associated variables and coefficients, including radiation energy $E_{r}$, radiation pressure $p_{r}$, thermal conduction coefficient $\kappa_{g}$, radiative energy diffusion coefficient $\chi_{r,g}$, and radiative energy exchange rate $\omega_{r,g}$, are set to zero. Regarding the EOSs, the water is described by the SG model (parameters are listed in Table~\ref{tab:SGEOS_Para}), while the gas phase is treated as an ideal gas with $\gamma_{g}=1.4$.

The simulation is performed in 2D Cartesian coordinates. Initially a circular bubble is centered at $\left( 0,\ 0\right)\ \mathrm{mm}$ with a radius of $3\ \mathrm{mm}$. The density and pressure inside the bubble are $1\ \mathrm{kg}/\mathrm{m}^{3}$ and $10^{5}\ \mathrm{Pa}$, respectively. An incident shock wave is positioned at $5.4\ \mathrm{mm}$ to the left of the bubble center initially, and propagates to the $x+$ direction. The right (with the superscript $R$) and left (with the superscript $L$) sides of the shock are initialized as follows:
\begin{equation}
    \begin{cases}
    \rho^{R} = 1000\ \mathrm{kg}/\mathrm{m}^{3},\ p^{R} = 10^{5}\ \mathrm{Pa},\ u^{R} = 0; \\
    \rho^{L} = 1323.65\ \mathrm{kg}/\mathrm{m}^{3},\ p^{L} = 1.9\times10^{9}\ \mathrm{Pa},\ u^{L} = 681.58\ \mathrm{m}/\mathrm{s}.
    \end{cases}
\end{equation}
The computational domain spans $\left[ -15,\ 15\right]\times\left[ -15,\ 15\right]\ \mathrm{mm}$, and the bubble radius ($3\ \mathrm{mm}$) is initially discretized by $400$ grid points.

\begin{table}
    \centering
    \begin{tabular}{cccc}
    \qquad \qquad \qquad & \qquad $t_{c}\left ( \upmu\mathrm{s} \right )$ \qquad & \qquad $v_{j}\left ( \mathrm{m/s} \right )$ \qquad & \qquad $p_{w}\left ( \mathrm{GPa} \right )$ \qquad \\
    Present  & \qquad $3.71$ \qquad & \qquad $2844$ \qquad & \qquad $4.75$ \qquad \\
    \cite{lin2016}  & \qquad $3.70$ \qquad & \qquad $2832$ \qquad & \qquad $5.90$ \qquad \\
    \cite{hawker2012}  & \qquad $3.66$ \qquad & \qquad $2810$ \qquad & \qquad $5.89$ \qquad \\
    \cite{bo2014}  & \qquad $3.70$ \qquad & \qquad $2830$ \qquad & \qquad - \qquad \\
    \cite{nourgaliev2006}  & \qquad $3.69$ \qquad & \qquad $2850$ \qquad & \qquad $10.1$ \qquad \\
    \end{tabular}
    \caption{Shock-induced bubble collapse compared to prior simulations: collision time ($t_{c}$), jet speed at collision ($v_{j}$), and water-hammer shock pressure ($p_{w}$)}
    \label{tab:SBI_vali}
\end{table}

Figure~\ref{fig:ShockBubble} demonstrates different stages of the shock-induced bubble collapse in the simulation, visualized by pressure contours and numerical schlieren. Upon shock impact, a reflected rarefaction wave forms in the liquid and a transmitted shock develops inside the bubble, as shown in figure~\ref{fig:ShockBubble}($a$). At the same time, a re-entrant liquid jet forms at the left side of the bubble and propagates in the same direction as the incident shock, as shown in figure~\ref{fig:ShockBubble}($b$). Upon impact of the jet with the rear interface of the bubble, a water-hammer shock is generated (see e.g. $t=3.9\ \upmu \mathrm{s}$ in figure~\ref{fig:ShockBubble}($c$)). The characteristic parameters of this event include the collision time $t_c$, the water-hammer shock pressure $p_w$, and the jet speed $v_j$ at the instant of impact. Subsequently, the jet penetrates the bubble and merges with the surrounding liquid, while secondary jets gradually develop (see figure~\ref{fig:ShockBubble}($d$)). These flow features closely resemble the numerical results reported in~\cite{hawker2012}. A quantitative comparison on $t_c$, $v_j$, and $p_w$ with the literature is provided in Table~\ref{tab:SBI_vali}. Good agreement is achieved for $t_c$ and $v_j$, while $p_w$ is relatively lower in present simulation, primarily due to coarser spatial resolution. The results confirm that the proposed radiation two-phase flow model correctly reduces to a standard compressible two-phase flow formulation when the radiation-specific terms are disabled. Furthermore, the good agreement with expected flow dynamics validates the core flow solver and the implementation of the equations of state before the coupling with radiation transport and thermal conduction is introduced. 

\subsection{Nanosecond laser pulse irradiates tin droplet}
\label{subsec:laser-drop}
\begin{figure}
\begin{center}
\includegraphics[width=13.2cm]{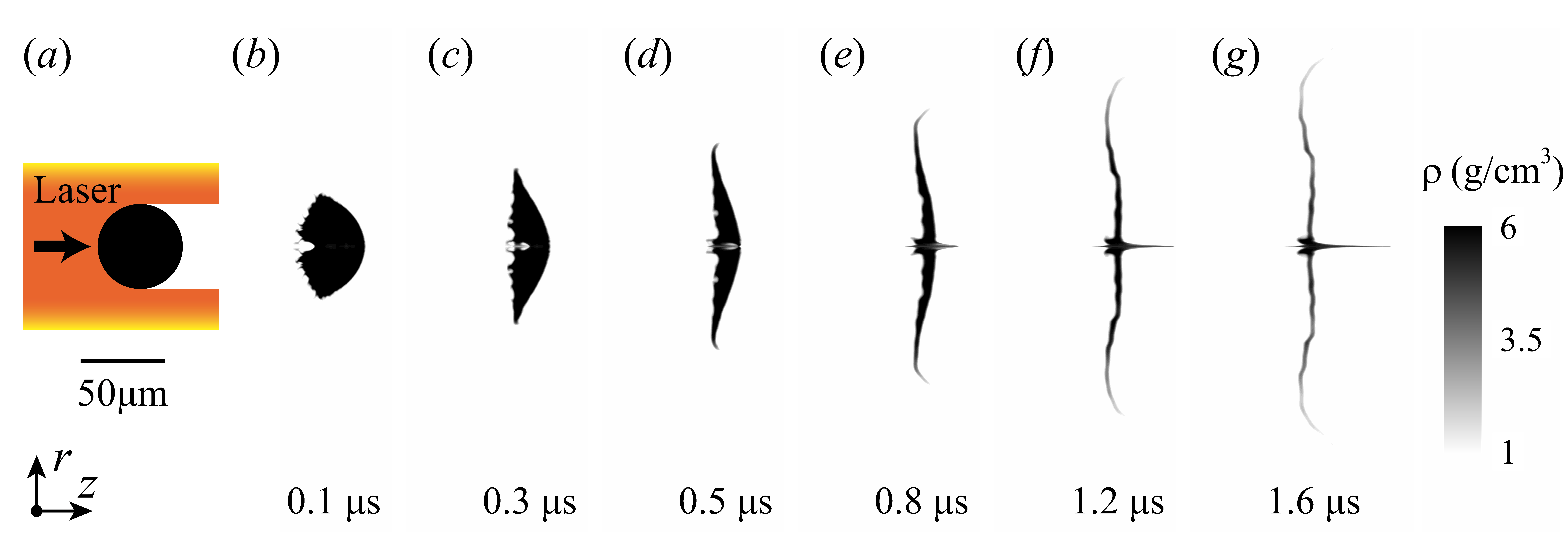}
\caption{($a$) Schematic of axisymmetric simulations of laser-droplet interactions. ($b$-$g$) The cross-sectional shadowgraph of the droplet at different times ranging from $t=0.1$ to $1.6\ \upmu\mathrm{s}$. All panels have the same spatial scale, with time labeled below each panel.}
\label{fig:LDsnapshots}
\end{center}
\end{figure}

Finally, we use the proposed model to simulate an experimentally relevant pre-pulse scenario in LPP-EUV lithography, in which a tin droplet is irradiated by a nanosecond laser pulse and deforms into a thin sheet. The simulation is configured as a 2D axisymmetric case with parameters matched to the systematic experimental work of~\cite{kurilovich2016}. The initial configuration is sketched in figure~\ref{fig:LDsnapshots}($a$). Specifically, a $7\ \mathrm{mJ}$ Nd:YAG laser pulse with the wavelength of $1064\ \mathrm{nm}$ is employed, with a duration of $10\ \mathrm{ns}$ full width at half maximum (FWHM) and the focal spot size of $115\ \upmu \mathrm{m}$ FWHM on the target. The laser is modeled by using a ray‑tracing approach, in which the beam is represented by over 20,000 individual rays that propagate geometrically through the plasma, each carrying a fraction of the total pulse energy. For this purpose, we adopt the ray‑tracing methodology detailed in our earlier work~\citep{tao2025}. Energy deposition in the plasma occurs primarily through inverse bremsstrahlung absorption, which is incorporated as a source term on the right-hand side of the mixture energy equation in Eq.~(\ref{eq:summary}). 

\begin{figure}
\centering
\includegraphics[width=13cm]{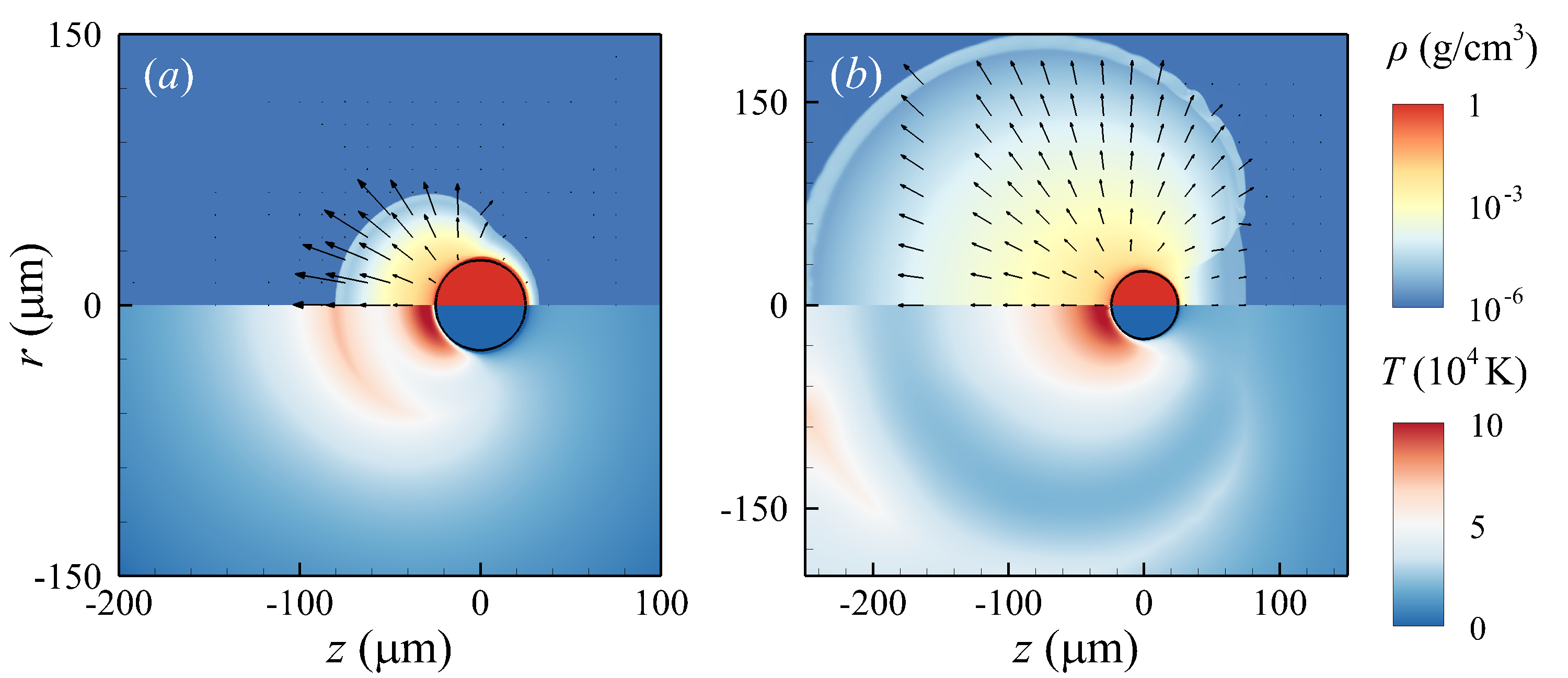}
\caption{Numerical results of laser-droplet interaction at $2\ \mathrm{ns}$ ($a$) and $10\ \mathrm{ns}$ ($b$), in terms of density (upper half) and temperature (lower half). The arrows denote velocity vectors and the black lines represent droplet shape (by $\alpha_{l}=0.5$).}
\label{fig:LD_10ns}
\end{figure}
A tin ($\rho_l = 6.92\ \mathrm{g/cm}^{3}$) droplet with a diameter of $D = 50\ \upmu\mathrm{m}$ is placed on the axis of the laser, with the ambient environment filled with a low-density tin plasma at $10^{-6}\ \mathrm{g}/\mathrm{cm}^{3}$ and $10\ \mathrm{Pa}$. To approximate the laser‑induced plasma formed by vaporization and ionization, which have not yet been self-consistently included in the proposed model, we prescribe a thin layer of high-density tin plasma ($1\ \upmu\mathrm{m}$ thick, with a density of $1\ \mathrm{g}/\mathrm{cm}^{3}$ and a pressure of $10^{7}\ \mathrm{Pa}$) on the left (laser‑irradiated) surface of the droplet. The mass in this layer is consistent with experimental observations, i.e. less than 1\% of the droplet mass is ablated during the laser‑plasma interaction stage~\citep{kurilovich2016}. The tin droplet is modeled as a stiffened gas with the parameters listed in Table~\ref{tab:SGEOS_Para}, while the tin plasma is treated as an ideal gas with $\gamma_{g}=5/3$ and $M_{g} = 118.71\ \mathrm{g/mol}$. The ionization degree ($Z_{g}$) is determined using the Thomas-Fermi model~\citep{more1985}, the thermal conduction coefficient ($\kappa_{g}$) follows the Spitzer-Harm theory~\citep{spitzer1953}, and the radiative transport coefficients ($\chi_{r,g}$, $\omega_{r,g}$) are evaluated using the empirical formulae from~\cite{tsakiris1987}; see details in Appendix~\ref{D}. Radiation is initially set to be in thermal equilibrium with the materials. The computational domain spans $\left[0, 300\right]\ \upmu \mathrm{m}$ in the radial direction ($r$) and $\left[-300, 300\right]\ \upmu \mathrm{m}$ in the axial direction ($z$), and initially $400$ grid points are used to resolve the droplet diameter.

\begin{figure}
\centering
\includegraphics[width=12.5cm]{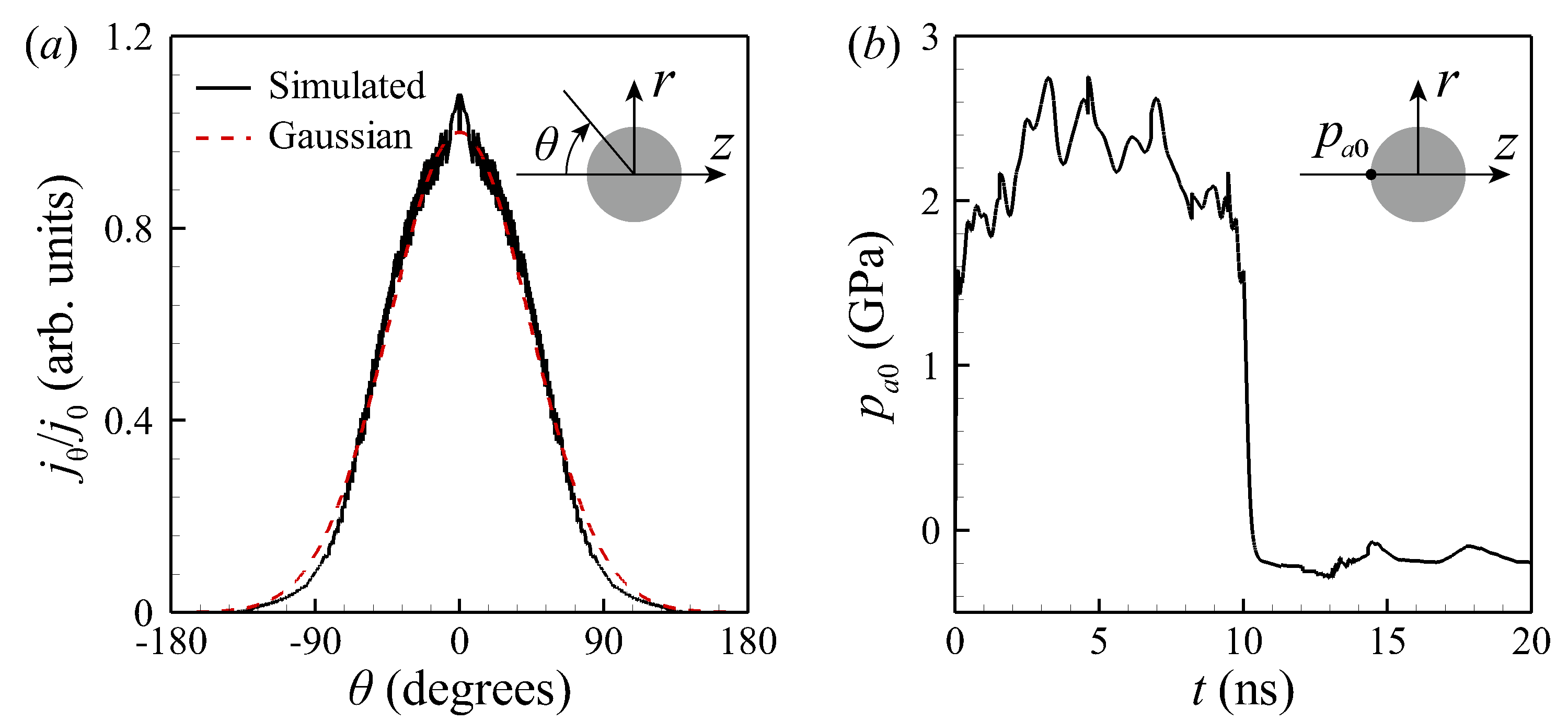}
\caption{($a$) Normalized surface impulse profiles: simulation results (black solid) and Gaussian fit (red dashed). The inset shows the polar angle $\theta$ in the $r$-$z$ coordinate. ($b$) Temporal evolution of the pressure exerted on the droplet surface at $\theta=0$, $p_{a0}$.}
\label{fig:LD_Psimu}
\end{figure}

Figure~\ref{fig:LD_10ns} shows the laser-induced plasma expansion at $t=2\ \mathrm{ns}$ and $10\ \mathrm{ns}$ during the laser-plasma interaction stage of the simulation. Under the left-side irradiation, the plasma expands rapidly, with the fastest motion occurring along the axial direction. The continuous laser energy deposition sustains a very high temperature ($\sim 10^{5}\ \mathrm{K}$) in the plasma near the irradiated surface of the droplet. The laser is absorbed below the critical density, and consequently forms a quasi-stationary ablation front, in consistence with the simulations by~\cite{basko2015}. In contrast, the droplet itself remains in a low-temperature, high-density liquid state, as illustrated by figure~\ref{fig:LD_10ns}. Since the laser-pulse duration ($10\ \mathrm{ns}$) is much shorter than the inertial timescale of droplet dynamic response ($\sim D/U \approx 1\ \upmu\mathrm{s}$, where $U$ is the axial propulsion velocity taken from Table~\ref{tab:LD_Vali}), the droplet shape has negligible deformation during the laser pulse. Significant deformation due to the high pressure at the ablation front is expected to occur on a longer timescale. 

\begin{figure}
\centering
\includegraphics[width=13cm]{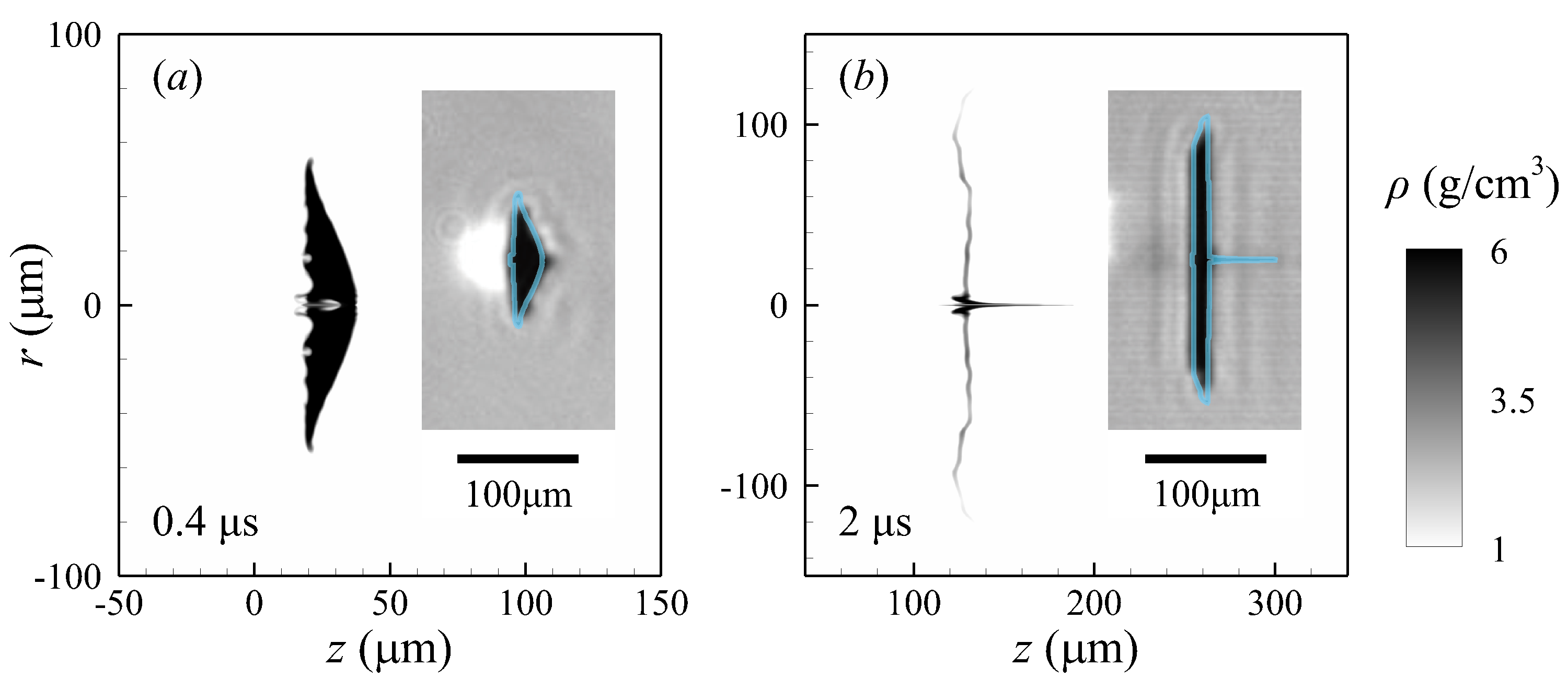}
\caption{Density contours of the tin sheet at $0.4\ \upmu \mathrm{s}$ ($a$) and $2\ \upmu \mathrm{s}$ ($b$). Insets show side-view experimental shadowgraphs (adapted from~\cite{kurilovich2016}), superimposed by the simulated drop shape (with respect to the contour of $\rho=1.2\ \mathrm{g/cm^{3}}$) representing the side-view projection.}
\label{fig:LD_2000ns}
\end{figure}

The in-flight deformation of the spherical droplet into a flat sheet over the inertial time scale is illustrated in figure~\ref{fig:LDsnapshots}($b$-$g$). The deformation starts on the laser-irradiated (left) surface, where fine structures develop (see figure~\ref{fig:LDsnapshots}$b$). Subsequently, the droplet assumes a jellyfish-hat-like shape (figure~\ref{fig:LDsnapshots}$c$), and is gradually flattened and curves away from the direction of the laser beam (figure~\ref{fig:LDsnapshots}$d$-\ref{fig:LDsnapshots}$g$). A notable feature during this flattening stage is the generation of an axial jet (figure~\ref{fig:LDsnapshots}$e$-\ref{fig:LDsnapshots}$g$), consistent with the experimental observations by~\cite{meijer2022}. It is remarkable that our proposed model successfully captures both this axial jet and the surficial fine structures, which have rarely been reproduced in previous incompressible or single-phase radiation hydrodynamic simulations. 

During the laser-plasma interaction stage, the expanding plasma exerts on the droplet surface a very high ablation pressure that decays rapidly after the laser being turned off, in agreement with previous single-phase simulations~\citep{kurilovich2018}. The surface impulse arising from the ablation pressure can be defined as $j_{\theta}=\int_{\tau} p_{a}\left(t,\ \theta\right)\mathrm{d}t$, where $p_{a}(t,\theta)$ is the surface pressure, $\theta$ is the polar angle shown in figure~\ref{fig:LD_Psimu}($a$), and the integration interval $\tau$ covers the laser pulse duration (in this case $\tau=10\ \mathrm{ns}$). This surface impulse model has been widely used~\citep{gelderblom2016,reijers2017,francca2025} to assess the subsequent droplet deformation in the flattening stage, e.g. by serving as an input for the incompressible droplet simulations. Figure~\ref{fig:LD_Psimu}($a$) displays the simulated profile of the normalized $j_{\theta}/j_{0}$, where $j_{0}=j_{\theta=0}$ is $21.9\ \mathrm{Pa \cdot s}$. We note that the profile is well fitted by a Gaussian function. The theoretical work in~\cite{francca2025} suggested that a Gaussian impulse profile may be insufficient to induce a curvature reversal, where the liquid sheet curves away from the laser side, within an incompressible droplet model. However, our simulation demonstrates such a reversal under an effectively Gaussian impulse (see figure~\ref{fig:LDsnapshots}$d$). Figure~\ref{fig:LD_Psimu}($b$) shows the temporal evolution of the surface pressure $p_{a}$ at $\theta =0$, $p_{a0}$. During the laser irradiation, the value of $p_{a0}$ ranges roughly from 1.5 GPa to 2.8 GPa, and its evolution exhibits multiple rises and drops, likely caused by pressure waves propagating across the ablation front.

Figure~\ref{fig:LD_2000ns} shows the deformed droplet at $t=0.4\ \upmu \mathrm{s}$ and $2\ \upmu \mathrm{s}$ in terms of density contours. For direct comparison, the insets superimpose the projection outlines, obtained by azimuthally rotating the simulated droplet profiles, onto the corresponding experimental results from~\cite{kurilovich2016}. Clearly, an excellent agreement has been achieved in the droplet shapes. Quantitative results, including the radial width $l_{r}$, axial thickness $d_{z}$, and axial propulsion velocity $U$ of the tin sheet at these two moments, are listed in Table~\ref{tab:LD_Vali} and compared with the experimental measurements. The results show reasonably good quantitative agreement, despite the simplified models of equations of state, thermal conduction, ionization, and radiation within the simulation. 
\begin{table}
    \centering
    \begin{tabular}{ccccc}
    \qquad \qquad \qquad & \qquad $t\left ( \upmu\mathrm{s} \right )$ \qquad & \qquad $l_{r}\left ( \upmu\mathrm{m} \right )$ \qquad & \qquad $d_{z}\left ( \upmu\mathrm{m} \right )$ \qquad & \qquad $U\left ( \mathrm{m/s} \right )$ \qquad \\
    Present & \qquad \multirow{2}{*}{0.4} \qquad  & \qquad 110 \qquad &  \qquad 26 \qquad & \qquad 58.81 \qquad \\
    \cite{kurilovich2016} &  & \qquad 100 \qquad & \qquad 30 \qquad & \qquad 54 \qquad \\
    Present & \qquad \multirow{2}{*}{2} \qquad & \qquad 220 \qquad &  \qquad 20 \qquad & \qquad  \qquad \\
    \cite{kurilovich2016} &  & \qquad 210 \qquad & \qquad 22 \qquad & \qquad  \qquad \\
    \end{tabular}
    \caption{Comparison with experimental observations for deformed tin droplet shape: radial width ($l_{r}$), axial thickness ($d_{z}$) and propulsion velocity ($U$) at $0.4\ \upmu\mathrm{s}$ and $2\ \upmu\mathrm{s}$.}
    \label{tab:LD_Vali}
\end{table}

\section{Conclusion}
\label{sec:conc}
We have developed a radiation two-phase flow model for plasma-liquid interactions using a diffuse interface framework. This approach approximates the physically sharp interface with a transition layer with finite thickness. Within this diffuse region, we formulated consistent energy flux closures that satisfy the correct physical jump conditions at the sharp interface limit, notably requiring the thermal and radiative energy fluxes from the plasma to vanish at the liquid surface. Additionally, appropriate equation-of-state mixing rules were introduced to ensure the physical propagation of acoustic waves throughout the mixture. The model was advanced in time using an operator-splitting numerical algorithm, sequentially handling hyperbolic convection and parabolic diffusion. After comprehensive validation against three test cases from the literature, the model showed excellent performance for both single-phase plasma flows and liquid-gas two-phase flows. We then employed the proposed model to simulate an LPP-EUV relevant pre-pulse scenario: A tin droplet is irradiated by a nanosecond laser pulse and deforms into a thin sheet. Our results show good agreement with the experimental findings of~\cite{kurilovich2016}. The simulation successfully captures key experimentally observed features such as the axial jet, which were rarely captured in previous simulations. Furthermore, it provides quantitatively accurate predictions of crucial dynamic properties, including the radial radius, axial thickness and axial propulsion velocity of the tin sheet.

By successfully simulating this complex LPP-EUV scenario, the proposed model establishes a framework for the self-consistent simulation of coupled laser-plasma physics and droplet dynamics. This framework is well-suited for incorporating additional essential physics, such as phase transition and more detailed material descriptions (e.g. tabulated equations of state), thereby enabling fundamental studies of plasma-liquid interactions critical for the optimization of state-of-the-art LPP-EUV sources.

\backsection[Funding]{We are grateful for the support of the National Natural Science Foundation of China (grant numbers: 12388101, 12375243), the Strategic Priority Research Program of Chinese Academy of Sciences (grant number: XDA0380601) and the Science Challenge Project (grant number: TZ2025016)
}

\backsection[Declaration of interests]{The authors report no conflict of interest.}

\appendix
\section{Governing equation for convection of the total energy density}
\label{A}
The governing equation for the total energy ($\rho e_{t}=\rho e+E_{r}+\rho \left|\mathbf{u}\right|^{2}/2$) can be obtained by summing mixture energy equation Eq.~(\ref{eq:Conv_part5}), radiation energy equation Eq.~(\ref{eq:Conv_part6}), and the dot product of momentum equation Eq.~(\ref{eq:Conv_part4}) with the velocity vector. Specifically, performing a dot product on both side of Eq.~(\ref{eq:Conv_part4}) with $\mathbf{u}$:
\begin{equation}
    \mathbf{u}\cdot\left[\partial_{t} \left ( \rho\mathbf{u} \right )  +\nabla \cdot \left ( \rho \mathbf{u}\mathbf{u} \right )\right]+\mathbf{u}\cdot\nabla p_{t} =0,
\end{equation}
by applying the identity:
\begin{equation}
    \begin{cases}
    \mathbf{u}\cdot\partial_{t} \left ( \rho\mathbf{u} \right ) =\partial_{t} \left ( \rho\left|\mathbf{u}\right|^{2}/2 \right ), \\
    \mathbf{u}\cdot\nabla \cdot \left ( \rho \mathbf{u}\mathbf{u} \right )=\nabla \cdot \left ( \rho\left|\mathbf{u}\right|^{2}\mathbf{u} /2\right ),\\
    \mathbf{u}\cdot\nabla p_{t}=\nabla\cdot \left(p_{t}\mathbf{u}\right)-p_{t}\nabla\cdot\mathbf{u},
    \end{cases}
\end{equation}
we can obtain:
\begin{equation}
    \partial_{t} \left ( \rho\left|\mathbf{u}\right|^{2}/2 \right )  +\nabla \cdot \left ( \rho\left|\mathbf{u}\right|^{2}\mathbf{u} /2\right )+\nabla\cdot \left(p_{t}\mathbf{u}\right)-p_{t}\nabla\cdot\mathbf{u} =0.
    \label{eq:MomentumA}
\end{equation}
Thus the sum of Eq.~(\ref{eq:Conv_part5}), Eq.~(\ref{eq:Conv_part6}), and Eq.~(\ref{eq:MomentumA}) leads to the convection equation of the total energy density as:
\begin{equation}
    \partial _{t} \left ( \rho e_{t}\right ) +\nabla \cdot \left ( \rho e_{t} \mathbf{u} \right ) + \nabla \cdot\left (  p_{t}\mathbf{u} \right) = 0 .
    \label{eq:totaleA}
\end{equation}

\section{Jacobian matrix and eigenvalues for the hyperbolic system}
\label{B}
The hyperbolic subsystem Eq.~(\ref{eq:Convection}), in one-dimensional cases, can be rewritten in terms of the conservative variables ${\bf U} = \left(\rho_l \alpha_l,\ \rho_g \alpha_g,\ \rho u,\ \rho e,\ E_r,\ \alpha_l\right)^\mathrm{T}$ as
\begin{equation}
    \partial_{t}\mathbf{U} + {\bf A}({\bf U}) \partial_{x}{\bf U} = 0.
\end{equation}
The Jacobian matrix $\mathbf{A}$ is written in following form:
\begin{equation}
	\notag
    \renewcommand{\arraystretch}{1.7}
	{\bf A} = 
	\begin{bmatrix}
        \displaystyle\frac{\rho_{g} \alpha_{g}u}{\rho} & \displaystyle-\frac{\rho_{l} \alpha_{l}u}{\rho} & \displaystyle\frac{\rho_{l} \alpha_{l}}{\rho} & 0 & 0 & 0\\
		\displaystyle-\frac{\rho_{g} \alpha_{g}u}{\rho} & \displaystyle\frac{\rho_{l} \alpha_{l}u}{\rho} & \displaystyle\frac{\rho_{g} \alpha_{g}}{\rho} & 0 & 0 & 0 \\
		-u^2 & -u^2 & 2u & \gamma-1 & \displaystyle\frac{1}{3} & W \\
		\displaystyle-\frac{(\rho e+p)u}{\rho} & \displaystyle-\frac{(\rho e+p)u}{\rho} & \displaystyle\frac{\rho e+p}{\rho} & u & 0 & 0\\
        \displaystyle-\frac{(E_r+p_r)u}{\rho} & \displaystyle-\frac{(E_r+p_r)u}{\rho} & \displaystyle\frac{E_r+p_r}{\rho} & 0 & u & 0\\
        0 & 0 & 0 & 0 & 0 & u
	\end{bmatrix},
\end{equation}
where $W = \partial p /\partial \alpha_{l}$. By applying the EOS of the mixture Eq.~(\ref{eq:mixture_EOS}), it can be written as:
\begin{equation}
	W=\frac{\left(\gamma-1\right)^{2}}{\left(\gamma_l-1\right)\left(\gamma_g-1\right)}\left(\left(\gamma_l-\gamma_g\right)\rho e+\gamma_g \mathrm{P}_{\infty,g}-\gamma_l \mathrm{P}_{\infty,l}\right).
\end{equation}
$\mathbf{A}$ possesses six real eigenvalues, i.e. $\mathbf{\lambda}=\left \{ u,\ u,\ u,\ u,\ u-C_{s},\ u+C_{s} \right \} $, where the characteristic wave speed $C_{s}$ of the system is expressed as:
\begin{equation}
    C_{s}=\sqrt{C^{2}+\frac{4p_{r}}{3\rho}},
\end{equation}
then the right eigenvectors of the matrix $\mathbf{A}$ are:\\
$\mathbf{R}_{1}\left(\mathbf{U}\right)=\left(0,\ 0,\ 0,\ -W,\ 0,\ \gamma-1\right)^{\mathrm{T}}$,\\
$\mathbf{R}_{2}\left(\mathbf{U}\right)=\left(0,\ 0,\ 0,\ -1,\ 3\left(\gamma-1\right),\ 0\right)^{\mathrm{T}}$,\\
$\mathbf{R}_{3}\left(\mathbf{U}\right)=\left(1,\ 0,\ u,\ 0,\ 0,\ 0\right)^{\mathrm{T}}$,\\
$\mathbf{R}_{4}\left(\mathbf{U}\right)=\left(-1 ,\ 1,\ 0,\ 0,\ 0,\ 0\right)^{\mathrm{T}}$,\\
$\mathbf{R}_{5}\left(\mathbf{U}\right)=\left(\rho_{l}\alpha_{l},\ \rho_{g}\alpha_{g},\ \rho \left(u-C_{s}\right),\ \rho C^{2}/\left(\gamma-1\right),\ 4p_{r},\ 0\right)^{\mathrm{T}}$,\\
$\mathbf{R}_{6}\left(\mathbf{U}\right)=\left(\rho_{l}\alpha_{l},\ \rho_{g}\alpha_{g},\ \rho \left(u+C_{s}\right),\ \rho C^{2}/\left(\gamma-1\right),\ 4p_{r},\ 0\right)^{\mathrm{T}}$.

\section{Calculation of the mixture internal energy and radiation energy from the total energy}
\label{C}
Advancing Eq.~(\ref{eq:totale}) from the $n$ time step to the intermediate state ($*$) would yield the total energy $(\rho e_{t})^{*}$. However, our target variables are the internal energy $(\rho e)^{*}$ and the radiation energy $(E_{r})^{*}$. The main reason why we do not directly advance Eq.~(\ref{eq:Conv_part5}) and Eq.~(\ref{eq:Conv_part6}) is that the calculation of the pressure work terms involving $\nabla \cdot \mathbf{u}$ becomes problematic in the presence of shocks. 

The discretized form of Eq.~(\ref{eq:Conv_part5}) and Eq.~(\ref{eq:Conv_part6}) indicates the energy increments due to advection (superscript $adv$) and pressure work (superscript $work$):
\begin{equation}
    (\rho e)^* - (\rho e)^n = \underbrace{- \nabla \cdot (\rho e \mathbf{u})^n \Delta t}_{\displaystyle\Delta (\rho e)^{adv}}\ \ \underbrace{- (p \nabla\cdot  \mathbf{u})^n \Delta t}_{\displaystyle\Delta (\rho e)^{work}},
    \label{eq:DeltarhoeB}
\end{equation}
\begin{equation}
    (E_{r})^* - (E_{r})^n = \underbrace{- \nabla \cdot (E_{r} \mathbf{u})^n \Delta t}_{\displaystyle\Delta (E_{r})^{adv}}\ \ \underbrace{- (p_{r} \nabla\cdot  \mathbf{u})^n \Delta t}_{\displaystyle\Delta (E_{r})^{work}}.
\label{eq:DeltarhoerB}
\end{equation}
Energy conservation requires that 
\begin{equation}
    \Delta \left(\rho e_{t}\right)=\Delta \left(\rho e_{k}\right)+\Delta \left(\rho e\right)^{adv}+\Delta \left(E_{r}\right)^{adv}+\Delta \left(\rho e\right)^{work}+\Delta \left(E_{r}\right)^{work},
    \label{eq:ConvELaw}
\end{equation}
where the increments on the total energy ($\Delta \left(\rho e_{t}\right)$) and the kinetic energy ($\Delta \left(\rho e_{k}\right)$) have been computed via
$$
\begin{cases}
    \Delta \left(\rho e_{t}\right)=\left(\rho e_{t}\right)^{*}-\left(\rho e_{t}\right)^{n}, \\
    \Delta \left(\rho e_{k}\right)=\left(\rho \left|\mathbf{u} \right|^{2}/2\right)^{*}-\left(\rho \left|\mathbf{u} \right|^{2}/2\right)^{n}.
\end{cases}
\label{eq:DeltaEtot}
$$
We then need to find the four unknowns $\Delta \left(\rho e\right)^{adv},\ \Delta \left(E_{r}\right)^{adv},\ \Delta \left(\rho e\right)^{work}$, and $ \Delta \left(E_{r}\right)^{work}$. 

The energy advections, i.e. $\Delta \left(\rho e\right)^{adv}$ and $\Delta \left(E_{r}\right)^{adv}$, are computed by being concurrently advanced with the hyperbolic subsystem (Eq.~(\ref{eq:Convection}$a$-$d$) and Eq.~(\ref{eq:totale})).
Then the pressure works ($\Delta \left(\rho e\right)^{work}$ and $\Delta \left(E_{r}\right)^{work}$) are determined by applying their ratio of $p_{r}/p$ from Eq.~(\ref{eq:DeltarhoeB}) and Eq.~(\ref{eq:DeltarhoerB}): 
\begin{equation}
    \begin{cases}
    \displaystyle\Delta \left(\rho e\right)^{work}=\frac{p}{p+p_{r}}\left(\Delta \left(\rho e_{t}\right)-\Delta \left(\rho e_{k}\right)-\Delta \left(\rho e\right)^{adv}-\Delta \left(E_{r}\right)^{adv}\right),\\
    \displaystyle\Delta \left(E_{r}\right)^{work}=\frac{p_{r}}{p+p_{r}}\left(\Delta \left(\rho e_{t}\right)-\Delta \left(\rho e_{k}\right)-\Delta \left(\rho e\right)^{adv}-\Delta \left(E_{r}\right)^{adv}\right).
    \end{cases}
    \label{eq:Work_P_Pr}
\end{equation}
Finally $\left(\rho e\right)^{*}$ and $\left(E_{r}\right)^{*}$ can be calculated by Eq.~(\ref{eq:DeltarhoeB}) and Eq.~(\ref{eq:DeltarhoerB}).

\section{Formulae for the physical models employed in the simulation of laser-droplet interaction}
\label{D}
In this paper, a series of physical models are used for simulating the pre-pulse scenario in LPP-EUV. In particular, the ionization degree ($Z_{g}$) is modeled by the Thomas-Fermi theory~\citep{more1985}, the thermal conduction coefficient ($\kappa_{g}$) is modeled by the Spitzer-Harm theory~\citep{spitzer1953}, and the radiative transport coefficients ($\chi_{r,g}$, $\omega_{r,g}$) are modeled by the empirical formulae from~\cite{tsakiris1987}. 
\subsection{Thomas-Fermi model}
\label{D1}
The Thomas-Fermi model~\citep{more1985} provides a useful approximation for the ionization degree $Z_{g}$ in high-density plasmas (e.g. LPP in EUV), specifically as:
\begin{equation}
    \displaystyle Z_{g} =  \frac{x}{1 + x + \sqrt{1 + 2x}}Z,
    \label{eq:TFmodel}
\end{equation}
in which
\begin{equation}
    \begin{cases}
    \displaystyle x=\eta \left( \frac{\rho_g}{AZ} \right)^\zeta \left[ 1 + \left( a_1 \left( \frac{T_g}{Z^{4/3}} \right)^{a_2} + a_3 \left( \frac{T_g}{Z^{4/3}} \right)^{a_4} \right)^L \left( \frac{\rho_g}{AZ} \right)^{(N-1)L} \right]^{\beta / L}, \\
    \displaystyle N = -\exp\left( b_0 + b_1 \left( \frac{T_g}{Z^{4/3} + T_g} \right) + b_2 \left( \frac{T_g}{Z^{4/3} + T_g} \right)^2 \right), \\
    \displaystyle L = c_1 \cdot \frac{T_g}{Z^{4/3} + T_g} + c_2.
    \end{cases}
\end{equation}
where the coefficients are $\eta=14.3139$, $\zeta=0.6624$, $a_{1}=0.003323$, $a_{2}=0.9718$, $a_{3}=9.26148\times10^{-5}$, $a_{4}=3.10165$, $b_{0}=-1.763$, $b_{1}=1.43175$, $b_{2}=0.31546$, $c_{1}=-0.366667$, and $c_{2}=0.983333$. $\rho_{g}$ and $T_{g}$ are the density and temperature of the plasma in the unit of $\mathrm{g/cm^{3}}$ and $\mathrm{eV}$, respectively. $A$ and $Z$ are the mass and atomic number of the element, respectively. Specifically,  $A = 118$ and $Z=50$ for tin. 

\subsection{Spitzer-Harm theory}
The Spitzer-Harm theory~\citep{spitzer1953} captures the dominant collisional energy exchange in high-temperature plasmas, ensuring local thermodynamic equilibrium. These conditions are approximately satisfied in the bulk region of our simulation domain:
\begin{equation}
    \kappa_{g}=\left(\frac{8}{\pi}\right)^{3/2}\frac{k_{B}^{7/2}}{e^{4}\sqrt{m_{e}}}\left ( \frac{1}{1+3.3/Z_{g}}  \right ) \frac{T_{g}^{5/2}}{Z_{g}\ln{\Lambda } } ,
    \label{eq:SpitzerHarm}
\end{equation}
where $k_{B}=1.3807\times10^{-16}\ \mathrm{g\cdot cm^{2}/\left(s^{2}\cdot K\right)}$ is the Boltzmann constant, $e=4.8032\times10^{-10}\ \mathrm{statcoulomb}$ is the electron charge, $m_{e}=9.1094\times10^{-28}\ \mathrm{g}$ is the mass of an electron, and $\ln{\Lambda }$ is the Coulomb logarithm given by:
\begin{equation}
    \ln{\Lambda }=\ln{\left[\sqrt{\frac{k_{B}T_{g}}{4\pi e^{2}n_{e}}}\bigg/\max\left(\frac{Z_{g}e^{2}}{3k_{B}T_{g}},\frac{\hbar}{2\sqrt{3k_{B}T_{g}m_{e}}}\right)\right]} ,
    \label{eq:Coulomb}
\end{equation}
where $n_{e}=\rho_{g}Z_{g}N_{A}/M_{g}$ is the electron number density (unit of $\mathrm{cm}^{-3}$, $N_{A}=6.022\times10^{23}\ \mathrm{mol}^{-1}$ is the Avogadro constant) and $\hbar=1.0546\times10^{-27}\ \mathrm{g\cdot cm^{2}/s}$ is the reduced Planck constant.

\subsection{Radiation empirical formulae}
The radiative transport coefficients ($\chi_{r,g}$, $\omega_{r,g}$) can be written in the following form:
\begin{equation}
    \begin{cases}
    \displaystyle\chi_{r,g} = \frac{ac}{3\rho_{g}\sigma_{R}}, \\
    \displaystyle\omega_{r,g} = ac\rho_{g}\sigma_{P},
    \end{cases}
    \label{eq:RossPlanck}
\end{equation}
where $a$ is the radiation constant and $c=3\times10^{10}\ \mathrm{cm/s}$ is the speed of light in vacuum. The Rossland $\sigma_{R}$ and Planck $\sigma_{P}$ mean opacities are given by the empirical formulae taken from~\cite{tsakiris1987}, which have been proven to be useful in the study of phenomena in hydrodynamics involving radiative energy transport (e.g. high-temperature high-density plasma):
\begin{equation}
    \begin{cases}
    \sigma_{R}[\mathrm{cm}^2\mathrm{/g}] = 72.19T_{g}^{-1.571}[\mathrm{keV}]\rho_{g}^{0.16}[\mathrm{g/cm}^{3}], \\
    \displaystyle\sigma_{P}[\mathrm{cm}^2\mathrm{/g}] = 328.55T_{g}^{-1.588}[\mathrm{keV}]\rho_{g}^{0.228}[\mathrm{g/cm}^{3}].
    \end{cases}
    \label{eq:Radempirical}
\end{equation}

\bibliographystyle{jfm}
\bibliography{reference}

\end{document}